\documentclass[twocolumn, 10pt, amsmath, amssymb]{revtex4}
\usepackage{epsfig, amssymb, stmaryrd}

\newcommand{\finpr}{\hfill $\square$ \vspace{2mm}}

\def\be{\begin{eqnarray}}
\def\ee{\end{eqnarray}}
\def\bee{\begin{eqnarray*}}
\def\eee{\end{eqnarray*}}
\newtheorem{thm}{Theorem}

\begin{document}

\title{Classical simulation versus universality in measurement based quantum computation}

\author{M. Van den Nest$^1$, W. D\"ur$^{1,2}$, G. Vidal$^3$, and H. J. Briegel$^{1,2}$ }

\affiliation{$^1$ Institut f\"ur Quantenoptik und Quanteninformation der \"Osterreichischen Akademie der Wissenschaften, Innsbruck, Austria\\
$^2$ Institut f{\"u}r Theoretische Physik, Universit{\"a}t
Innsbruck,
Technikerstra{\ss}e 25, A-6020 Innsbruck, Austria\\
$^3$ School of Physical Sciences, the University of
Queensland, QLD 4072, Australia.}
\date{\today}

\date{\today}
\def\makeheadbox{}

%\maketitle

\begin{abstract}
We investigate for which resource states an efficient
classical simulation of measurement based quantum
computation is possible. We show that the
\emph{Schmidt--rank width}, a measure recently introduced
to assess universality of resource states, plays a crucial
role in also this context. We relate Schmidt--rank width to
the optimal description of states in terms of tree tensor
networks and show that an efficient classical simulation of
measurement based quantum computation is possible for all
states with logarithmically bounded Schmidt--rank width
(with respect to the system size). For graph states where
the Schmidt--rank width scales in this way, we efficiently
construct the optimal tree tensor network descriptions, and
provide several examples. We highlight parallels in the
efficient description of complex systems in quantum
information theory and graph theory.
\end{abstract}

%\pacs{}

\maketitle

\section{Introduction}

The classical description of many--body quantum systems,
and the classical simulation of their dynamics, is
generically a hard problem, due to the exponential size of
the associated Hilbert space \cite{Fe85, Ek98}.
Nevertheless, under certain conditions an efficient
description of states and/or their evolution is possible.
This is, for instance, demonstrated by the density matrix
renormalization group method \cite{Sc04}, which allows one
to successfully calculate ground states of strongly
correlated spin systems in one spatial dimension using
matrix product states \cite{Ve05}. In this context, the
questions \emph{'For which (families of) states does an
efficient classical description exist?'}, and \emph{'When
is an efficient classical simulation of the evolution of
such states under a given dynamics possible?'} are
naturally of central importance.

Apart from their practical importance, the above questions
are directly related to more fundamental issues, in
particular to the power of quantum computation and the
identification of the essential properties that give
quantum computers their additional power over classical
devices; this relation to quantum computation will be
central in this article. In particular, we will study these
questions from the point of view of the \emph{measurement
based} approach to quantum computing, more specifically the
model of the one--way quantum computer \cite{Ra01}. In this
model, a highly entangled multi--qubit state, the 2D
\emph{cluster state} \cite{Br01}, is processed by
performing sequences of adaptive single--qubit
measurements, thereby realizing arbitrary quantum
computations. The 2D cluster state serves as a
\emph{universal resource} for measurement based quantum
computation (MQC), in the sense that any multi--qubit state
can be prepared by performing sequences of local operations
on a sufficiently large 2D cluster state.

When studying the fundamentals of the one--way model, two
(related) questions naturally arise, which we will consider
in the following; first, it is asked which resource states,
other than the 2D cluster states, form universal resources
for MQC; second, one may also consider the question whether
MQC on a given state can be \emph{efficiently simulated} on
a classical computer. Naturally, these two issues are
closely related, as one expects that an efficient classical
simulation of MQC performed on (efficient) universal
resource states is impossible. However, it is important to
stress that classical simulation and non--universality are
principally different issues.

The question of which other resource states are also
universal has been investigated recently in Ref.
\cite{Va06}, where the required entanglement resources
enabling universality were investigated.  In particular, it
was proven that certain entanglement measures, in
particular certain \emph{entanglement width} measures, must
diverge on any universal resource, thus providing necessary
conditions for universality.

On the other hand, the issue of classical simulation of MQC
evidently brings us back to the central introductory
questions posed above. Results regarding the efficient
simulation of MQC do exist, and it is e.g. known that any
MQC implemented on a 1D cluster state can be simulated
efficiently \cite{Ni05}. More generally, the efficient
description of quantum states in terms of (tree) tensor
networks turns out to play an important role in this
context \cite{Sh05, Sh'05} .

In this article we strengthen the connection between
classical simulation of MQC and non--universality. Our
starting point will be the no--go results for universality
obtained in Ref. \cite{Va06}, stating that the entanglement
monotones \emph{entropic entanglement width} and
\emph{Schmidt--rank width} must diverge on any universal
resource; both measures are closely related, and we refer
to section \ref{sect_Ewd} for definitions. We then focus on
the Schmidt--rank width measure, and prove, as our first
main result, that MQC can be efficiently simulated on every
resource state which is ruled out by the above no--go
result. More generally, we prove that MQC can be simulated
efficiently on all states where the Schmidt--rank width
grows at most logarithmically with the system size
\cite{Foot1}.

Second, along the way of proving the above results, we
provide a natural interpretation of the Schmidt--rank width
measure, as we show that this monotone quantifies what the
optimal description of quantum states is in terms of tree
tensor networks; this shows that there is in fact a large
overlap between the present research and the work performed
in Ref. \cite{Sh'05} regarding the simulation of quantum
systems using tree tensor networks.

As our third main result, we show that the Schmidt--rank
width (and entanglement width) -- these are measures which
are defined in terms of nontrivial optimization problems --
can be computed efficiently for all graph states. Moreover,
for all graph states where the Schmidt--rank width grows at
most logarithmically with the number of qubits, we give
efficient constructions of the optimal tree tensor networks
describing these states.

We further remark that the origin of the Schmidt-rank width
lies in fact in graph theory, and its definition is
inspired by a graph invariant called \emph{rank width}. It
turns out that the study of rank width in graph theory
shows strong similarities with the study of efficient
descriptions and simulations of quantum systems, viz. the
two introductory questions of this article. The similarity
is due to the fact that, in certain aspects of both quantum
information theory and graph theory, one is concerned with
the efficient description of complex structures in terms of
tree--like structures. We will comment on the existing
parallels between these fields.

Finally, we emphasize that the present work is situated in
two different dynamic areas of research within the field of
quantum information theory; the first is the study of
universality and classical simulation of measurement based
quantum computation, and the second is the problem of
efficiently describing quantum systems and their dynamics.
An important aim of this article consists of bringing
together existing results in both fields and showing that
there is a strong connection between them; in particular,
we find that the notion of Schmidt--rank width has been
considered independently in Refs. \cite{Va06} and
\cite{Sh'05} and plays an important role in both areas of
research. In order to establish the connections between
these two areas in a transparent manner, a substantial part
of this article is devoted to giving a clear overview of
which relevant results are known in both fields.

The paper is organized as follows. In section
\ref{universality} we discuss entanglement width and
Schmidt-rank width, and their role in universality and
classical simulation of MQC. In Section \ref{sect_tensor}
the description of states in terms of tree tensor networks
is reviewed, and a connection to Schmidt-rank width is
established. This Section also includes our main result,
stating that any state with a logarithmically bounded
Schmidt-rank width has, in principle, an efficient
description in terms of a tree tensor network, and hence
any MQC performed on such states can be efficiently
simulated classically. In Section \ref{sect_grapstates}
these results are applied to graph states, and we provide
in addition an explicit way of obtaining the optimal tree
tensor network. We discuss the relation between the
treatment of complex systems in quantum information theory
and graph theory in section \ref{sect_complex}, and
summarize and conclude in section \ref{sect_conclusion}.

%-------------------------------------------------------------------------------
%-------------------------------------------------------------------------------
\section{Entanglement width, universality, and classical simulation}
\label{universality}
%-------------------------------------------------------------------------------
%-------------------------------------------------------------------------------
%define ewd + swd
%review universality definition + results
%define efficient computation - scaling with N
%pose open questions

In this section we introduce two related multipartite
entanglement measures called \emph{entropic entanglement
width} and \emph{Schmidt--rank width} and discuss their
role in the studies of universality of resources for
measurement based quantum computation (MQC) and in
classical simulation of MQC.

These entanglement measures are defined in section
\ref{sect_Ewd}. In section \ref{sect_uni} we review the
definition of universal resources for MQC, and the use of
the above measures in this study. In section \ref{sect_sim}
we consider the basic notions regarding efficient classical
simulation of MQC. Finally, in section \ref{sect_problem}
we pose the two central questions of this article in a
precise way; the first question asks about the
interpretation of the measures entanglement width and
Schmidt--rank width, and the second deals with the role of
these measures in the context of classical simulation of
MQC.

\subsection{Entanglement width}\label{sect_Ewd}

The entropic entanglement width $E_{\mbox{\scriptsize
wd}}(|\psi\rangle)$ of an multi--party state $|\psi\rangle$
is an entanglement measure introduced in Ref. \cite{Va06}.
Qualitatively, this measure computes the minimal bipartite
entanglement entropy in the state $|\psi\rangle$, where the
minimum is taken over specific classes of bipartitions of
the system. The precise definition is the following.

\begin{figure}[ht]
\begin{picture}(230,160)
%\put(-5,0){\epsfxsize=230pt\epsffile[2 447 577
%759]{subcubic.eps}}
\put(0,0){\includegraphics[width=0.45\textwidth]{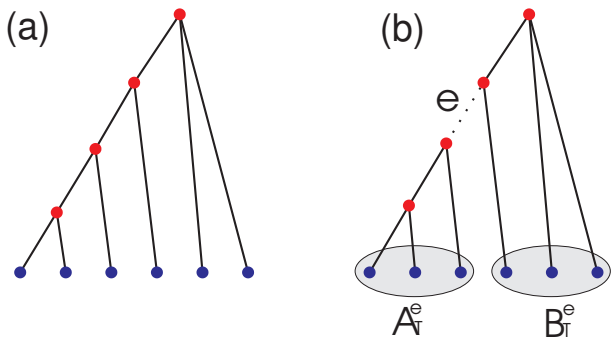}}
\end{picture}
\caption[]{\label{subcubic} (a) Example  of a subcubic tree
$T$ with six leaves (indicated in blue). (b) Tree
$T\setminus e$ obtained by removing edge $e$ and induced
bipartition $(A_{T}^e, B_{T}^e)$.}
\end{figure}

Let $|\psi\rangle$ be an $n$-party pure state. A
\emph{tree} is a graph with no cycles. Let $T$ be a
\emph{subcubic} tree, which is a tree such that every
vertex has exactly 1 or 3 incident edges. The vertices
which are incident with exactly one edge are called the
\emph{leaves} of the tree. We consider trees $T$ with
exactly $n$ leaves $V:=\{1, \dots, n\}$, which are
identified with the $n$ local Hilbert spaces of the system.
Letting $e=\{i, j\}$ be an arbitrary edge of $T$, we denote
by $T\setminus e$ the graph obtained by deleting the edge
$e$ from $T$. The graph $T\setminus e$ then consists of
exactly two connected components (see Fig. \ref{subcubic}),
which naturally induce a bipartition $(A_{T}^e, B_{T}^e)$
of the set $V$. We denote the bipartite entanglement
entropy of $|\psi\rangle$ with respect to the bipartition
$(A_{T}^e, B_{T}^e)$ by $E_{A_{T}^e,
B_{T}^e}(|\psi\rangle)$. The entropic entanglement width of
the state $|\psi\rangle$ is now defined by \be
E_{\mbox{\scriptsize wd}}(|\psi\rangle):= \min_T\
\max_{e\in T}\ E_{A_{T}^e, B_{T}^e}(|\psi\rangle), \ee
where the minimization is taken over all subcubic trees $T$
with $n$ leaves, which are identified with the $n$ parties
in the system.

Thus, for a given tree $T$ we consider the maximum, over
all edges in $T$, of the quantity $\ E_{A_{T}^e,
B_{T}^e}(|\psi\rangle)$; then the minimum, over all
subcubic trees $T$,  of such maxima is computed.

Similarly, one may use the Schmidt rank, i.e. the number of
non--zero Schmidt coefficients, instead of the bipartite
entropy of entanglement as basic measure. One then obtains
the \emph{Schmidt--rank width}, or \emph{$\chi$--width},
denoted by $\chi_{\mbox{\scriptsize{wd}}}(|\psi\rangle)$.
The precise definition is the following. Letting
$\chi_{A^e_T, B^e_T}(|\psi\rangle)$ denote the number of
non--zero Schmidt coefficients of $|\psi\rangle$ with
respect to a bipartition $(A_{T}^e, B_{T}^e)$ of $V$ as
defined above, the $\chi$--width of the state
$|\psi\rangle$ is defined by \be \label{chiwidth}
\chi_{\mbox{\scriptsize{wd}}} (|\psi\rangle):=
\min_{T}\max_{e\in T} \log_2 \chi_{A^e_T,
B^e_T}(|\psi\rangle). \ee

It is straightforward to show (cf. Ref. \cite{Va06}) that
$E_{\mbox{\scriptsize wd}}$ is an entanglement monotone
\cite{Vi98}, i.e., this measure vanishes on product states,
is a local invariant, and decreases on average under local
operations and classical communication (LOCC). The proof
can readily be extended to $\chi_{\mbox{\scriptsize wd}}$,
demonstrating that also $\chi$--width is a valid
entanglement measure. In fact, using that the Schmidt rank
is non--increasing under \emph{stochastic} LOCC, or SLOCC,
it can be proven that the $\chi$--width is also
non--increasing under SLOCC.

Since the inequality \be\log_2 \chi_{A,
B}(|\psi\rangle)\geq E_{A, B}(|\psi\rangle)\ee holds for
any bipartition $(A, B)$ of the system and for any state
$|\psi\rangle$, we have \be
\chi_{\mbox{\scriptsize{wd}}}(|\psi\rangle)  \geq
E_{\mbox{\scriptsize wd}} (|\psi\rangle).\ee Note, however,
that these quantities can show a completely different
(scaling) behavior.

It is clear that the definitions of entropic entanglement
width and Schmidt--rank width are based upon  similar
constructions, where optimizations are performed over
subcubic trees. Such constructions can of course be
repeated for any bipartite entanglement measure; hence a
whole class of multipartite entanglement measures is
obtained, which we will call the class of
\emph{entanglement width measures}. The entropic
entanglement width and $\chi$--width are two examples of
entanglement width measures. It would be interesting to
consider other examples of entanglement width measures, and
investigate their possible role in quantum information
theory.

The definitions of the  above entanglement measures are
inspired by a graph invariant called \emph{rank width},
which was introduced in Ref. \cite{Oum}. The connection
with rank width is obtained by evaluating the entropic
entanglement width or $\chi$--width in \emph{graph states}.
This is explained next.

First we  recall the definition of graph states. Let
$\sigma_x, \sigma_y, \sigma_z$ denote the Pauli spin
matrices. Let $G=(V, E)$ be a graph with vertex set
$V:=\{1, \dots, n\}$ and edge set $E$. For every vertex
$a\in V$, the set $N(a)$ denotes the set of neighbors of
$a$, i.e., the collection of all vertices $b$ which are
connected to $a$ by an edge $\{a, b\}\in E$.  The graph
state $|G\rangle$ is then defined to be the unique
$n$-qubit state which is the joint eigenstate, with
eigenvalues equal to 1, of the $n$ commuting correlation
operators \be\label{K_a} K_a:=
\sigma_x^{(a)}\bigotimes_{b\in N(a)} \sigma_z^{(b)}.\ee
Standard examples of graph states include the GHZ states,
and the 1D and 2D cluster states, which are obtained if the
underlying graph is a 1D chain or a rectangular 2D grid,
respectively. We refer to Ref. \cite{He06} for further
details.

Let $\Gamma$ be  the $n\times n$ adjacency matrix of $G$,
i.e, one has $\Gamma_{ab}=1$ if $\{a, b\}\in E$ and
$\Gamma_{ab}=0$ otherwise. For every bipartition $(A, B)$
of the vertex set $V$, define $\Gamma( A, B)$ to be the
$|A|\times |B|$ submatrix of $\Gamma$ defined by \be
\Gamma( A, B) := (\Gamma_{ab})_{a\in A, b\in B}.\ee Using
standard graph state techniques it can then be shown (see
e.g. \cite{He06}) that \be \label{ctrk} \mbox{
rank}_{\mathbb{F}_2}\ \Gamma( A, B)&=& \log_2 \chi_{A,
B}(|G\rangle)\nonumber\\&=& E_{A, B}(|G\rangle).\ee where
$\mbox{rank}_{\mathbb{F}_2}X$ denotes the rank of a matrix
$X$ when arithmetic is performed over the finite field
$\mathbb{F}_2=$ GF(2). Thus, the Schmidt rank and the
bipartite entanglement entropy w.r.t. any bipartition
$(A,B)$ coincide for graph states, and are given by the
rank of the matrix $\Gamma(A, B)$. Using the identity
(\ref{ctrk}), one immediately finds that the $\chi$--width
(and entropic entanglement width) of the graph state
$|G\rangle$ coincides with the \emph{rank width} rwd$(G)$
of the graph $G$. The explicit definition of rwd$(G)$ reads
\cite{Oum} \be\mbox{rwd}(G):= \min_{T}\max_{e\in T} \mbox{
rank}_{\mathbb{F}_2} \Gamma( A_T^e, B_T^e)\ee (where the
minimization is again over subcubic trees as in the
definition of $\chi$--width), which, using (\ref{ctrk}),
indeed coincides with the $\chi$--width of $|G\rangle$.

Note that the subcubic trees which are considered in the
definition of rank width are not to be confused with the
defining graph $G$ of the graph state $|G\rangle$ (the
latter can be an arbitrary graph); the subcubic trees
merely serve as a means of selecting certain bipartitions
of the system, independent of the state which is
considered. For instance, if we consider a linear  cluster
state $|L_6\rangle$ of six qubits, corresponding to a graph
$L_6$ that is a linear chain, then the tree depicted in
Fig. \ref{subcubic} corresponds to the optimal tree in the
definition of the rank-width (and $\chi$-width), leading to
rwd$(L_6)=\chi_{\rm wd}(|L_6\rangle)=1$.

In section \ref{sect_complex} we will further comment on
the motivations for the definition of rank width, and we
will draw parallels with the study of complex systems in
quantum information theory.

\subsection{Universal resources for MQC}\label{sect_uni}

In Ref. \cite{Va06} a definition for universality of
families of states for MQC was put forward, and the use of
$E_{\mbox{\scriptsize wd}}$ to assess non--universality of
states was demonstrated. In this section we briefly review
the definition and the corresponding results.

Consider an (infinitely large) family of qubit states
\be\Psi = \{|\psi_1\rangle,|\psi_2\rangle, \ldots\},\ee
where $|\psi_i\rangle$ is a state on $m_i$ qubits and $m_i<
m_{i+1}$ for every $i=1, 2, \dots$. This family is called a
\emph{universal resource for MQC} if for each  state
$|\phi\rangle$ on $n$ qubits there exists a state
$|\psi_i\rangle \in \Psi$ on $m_i$ qubits, with $m_i\geq
n$, such that the transformation $|\psi_i\rangle \to
|\phi\rangle|0\rangle^{m_i-n}$
 is possible deterministically by means of LOCC. That is, any state
$|\phi\rangle$ can be prepared using only states within the
family $\Psi$ as resource. Equivalently, the action of an
arbitrary unitary operation $U$ on a product input state
$|0\rangle^{n}$ can be implemented, where now $|\phi\rangle
:= U|0\rangle^{n}$ in the above definition. This definition
is in the spirit of the model of the one--way quantum
computer, where sequences of adaptive single--qubit
measurements performed on a sufficiently large 2D cluster
state allow one to prepare any multi--qubit state. The
definition of universal resource aims to identify the
required resources, in terms of entanglement, that allow
one to perform universal quantum computation in the sense
specified above.

In the above definition of universality of a family $\Psi$,
we have not yet considered the efficiency with which states
can be prepared using members of $\Psi$. An
\emph{efficient} universal resource $\Psi$ is a universal
resource having the property that all states that can be
efficiently generated with a quantum gate network should
also be efficiently generated from universal resource
$\Psi$. We refer to Ref. \cite{Va06b} for a detailed
account on efficient universality.

In Ref. \cite{Va06} it was found that any universal
resource $\Psi$ must satisfy the following property. Let
$E(|\phi\rangle)$ be a functional which is defined on the
set of \emph{all} $n$-qubit states, for \emph{all}
$n\in\mathbb{N}$, and suppose that $E(|\phi\rangle)$ is
non--increasing under LOCC. More precisely, if
$|\phi\rangle$ and $|\phi'\rangle$ are states on $n$ and
$n'$ qubits, respectively, then $E(|\phi\rangle) \geq
E(|\phi'\rangle)$ whenever the transformation $|\phi\rangle
\to |\phi'\rangle|0\rangle^{n-n'}$ is possible by means of
LOCC. Moreover, let $E^*$ denote the supremal value of
$E(|\phi\rangle)$, when the supremum is taken over all
$n$-qubit states, for all $n\in\mathbb{N}$ (the case
$E^*=\infty$ is allowed). Then any universal resource
$\Psi$ must satisfy the property \be \sup\
\{E(|\psi\rangle)\ |\ |\psi\rangle\in\Psi\} = E^*.\ee That
is, the supremal value of every entanglement measure $E$
must be reached on every universal resource $\Psi$. Using
the fact that there exist families of quantum states where
the entropic entanglement width and $\chi$--width grow
unboundedly with the system size (the 2D cluster states are
such examples), it is then straightforward to show that any
universal family of states ${\Psi}$ must have unbounded
entropic entanglement width and $\chi$--width as well. More
precisely, one has \cite{Va06}:
\begin{thm}\label{thm_ewd}
Let $\Psi$ be a universal resource for MQC. Then the
following statements hold:
\begin{itemize}
\item[(i)] $ \sup\ \{E_{\mbox{\scriptsize{wd}}}(|\psi\rangle)\ |\ |\psi\rangle\in\Psi\} = \infty$;
\item[(ii)] $ \sup\ \{\chi_{\mbox{\scriptsize{wd}}}(|\psi\rangle)\ |\ |\psi\rangle\in\Psi\} = \infty$.
\end{itemize}
\end{thm}
In other words, families $\Psi$ where the measures
$E_{\mbox{\scriptsize{wd}}}$  or
$\chi_{\mbox{\scriptsize{wd}}}$ are \emph{bounded}, cannot
be universal. This insight, together with the relation
between entropic entanglement width and $\chi$--width and
the graph theoretical measure rank width, allows one to
identify classes of graph states as being non--universal
since the rank width is bounded on such classes. Examples
include linear cluster graphs, trees, cycle graphs,
cographs, graphs locally equivalent to trees, graphs of
bounded tree--width, graphs of bounded clique--width or
distance--hereditary graphs. We refer to the literature for
definitions.

In the remainder of this paper, we will focus on the
$\chi$--width measure.

\subsection{Classical simulation of MQC}\label{sect_sim}

Rather than considering the question whether a family
$\Psi$ is a universal resource for MQC, one may also
consider the question whether MQC on $\Psi$ can be
\emph{efficiently simulated} on a classical computer. We
will say that efficient classical simulation of MQC on a
family of states $\Psi$ is possible, if for every state
$|\psi_i\rangle \in \Psi$ it is possible to simulate every
LOCC protocol on a classical computer with overhead
poly$(m_i)$, where $m_i$ denotes the number of qubits on
which the state $|\psi_i\rangle$ is defined, as before. We
remark that an efficient classical description of the
initial states $|\psi_i\rangle$ is a necessary, but not
necessarily a sufficient condition for efficient simulation
on a classical computer.

The issue of classical simulation  of MQC has recently been
considered by several authors. At this point we remind the
reader of what is already known in this context. Regarding
simulation of MQC on \emph{graph states}, we recall the
following results:
\begin{itemize}
\item In Ref. \cite{Ni05} it was showed that MQC on 1D
cluster states can be simulated efficiently classically;
\item In Ref. \cite{Sh'05} it was showed that MQC on tree
graphs can be simulated efficiently classically; \item In
Ref. \cite{Sh05} it was showed that MQC on graphs with
logarithmically bounded tree width \cite{Foot3} can be
simulated efficiently classically.
 \end{itemize}
Note that the above result on tree width implies the two
other results, as tree graphs (and thus also 1D cluster
graphs) have tree width equal to 1 \cite{Ro84}.

 More general results,
i.e., regarding \emph{arbitrary states}, were obtained in
Ref. \cite{Sh'05}, where it was shown that MQC can be
simulated efficiently on all states allowing an efficient
\emph{tree tensor network} description. The description of
quantum systems in terms of tree tensor networks will play
an important role in the present analysis, and will be
reviewed in detail in section \ref{sect_tensor}.

Although related, the issues of universality and classical
simulation in MQC are fundamentally two different
questions. Most of us expect that any family $\Psi$ for
which classical simulation of MQC is possible, will not be
an efficient universal resource; this reflects the common
belief that quantum computers are in some sense
exponentially more powerful than classical machines --
note, however, that so far there is no rigorous proof of
this statement. While one expects the possibility of
classical simulation of MQC to imply non--universality of a
resource $\Psi$, the converse implication is certainly not
believed to hold in general. Indeed, it is highly likely
that many non--universal families could still be used to
implement specific quantum algorithms.

\subsection{Problem formulation}\label{sect_problem}

It is clear that regarding the notion of $\chi$-width, and the above issues of universality and classical simulation of MQC, a number of open questions remain. In this section we formulate two central questions, (Q1) and (Q2), which will constitute the main research topics in this article. We will first state these questions and then discuss them.
\begin{itemize}
\item[(Q1)] Does there exist a natural
\emph{interpretation} of the $\chi$--width measure?
\item[(Q2)] Do there exist resources $\Psi$ having bounded
$\chi$--width, which nevertheless \emph{do not allow} an
efficient classical simulation of MQC?
\end{itemize}
Question (Q1) is concerned with the fact that the
definition of $\chi$--width seems to be rather arbitrary
and not intuitive, and solely motivated by the connection
to the graph theoretical measure rank width. We will,
however, provide a satisfactory  interpretation of this
measure in the context of quantum information in the next
section.

Question (Q2) is concerned with the question whether
non--universal resources can still be useful for quantum
computation, in the sense that MQC performed on such states
is more powerful than classical computation. As remarked
above, it may well be that there exist non--universal
families of states where MQC is nevertheless hard to
simulate classically. Previous results leave open this
possibility, as the criteria for non--universality and
classical simulatability do not coincide. For
non--universal states detected by the $\chi$--width
criterion (i.e., theorem \ref{thm_ewd} (ii)), we will show
that this is not the case. In section \ref{sect_connection}
we will show that MQC can be simulated efficiently for any
family $\Psi$ which is ruled out by the $\chi$--width
criterion as being a non--universal resource.

\section{Entanglement--width and Tree tensor
networks}\label{sect_tensor}

In this section we tackle questions (Q1) and (Q2) as stated
in the previous section.  First we will attach a natural
interpretation to the $\chi$--width measure, as  we will
show that $\chi_{\mbox{\scriptsize{wd}}}(|\psi\rangle)$
quantifies the complexity of the optimal \emph{tree tensor
network} (TTN) describing the state $|\psi\rangle$, thus
providing a satisfactory answer to question (Q1). Moreover,
we shall see that this connection with tree tensor networks
immediately allows us to give a negative answer to (Q2): we
find that MQC can be simulated efficiently on all resources
having a bounded $\chi$--width.

These results will be obtained in three main steps. In
section \ref{sectionTTN} we review the notions of tensor
networks and, more particularly, tree tensor networks. We
also review results  obtained in Ref. \cite{Sh'05}, where
it was proved that LOCC on states specified in terms of
efficient TTN descriptions can be simulated efficiently;
the results in Ref. \cite{Sh'05} will be central
ingredients to our analysis. In section
\ref{section_simTTN} we show how to obtain TTN descriptions
for arbitrary quantum states. Finally, in section
\ref{sect_connection} we establish the connection between
TTNs and $\chi$--width.

\subsection{Tree tensor networks and efficient simulation of quantum systems}\label{sectionTTN}
In this section we review the basic notions regarding
(tree) tensor networks (see also Ref. \cite{Sh05}), and the
simulation of quantum systems described by TTNs as obtained
in Ref. \cite{Sh'05}.

Consider a $d_1\times\dots\times d_n$ complex tensor
\cite{Foot4} \be A:=A_{i_1i_2\dots i_n},\ee where each
index $i_{\alpha}$ ranges from $1$ to $d_{\alpha}$, for
every $\alpha=1, \dots, n$. The number of indices $n$ is
sometimes called the \emph{rank} of the tensor $A$. We will
call the number $D:=\max_{\alpha}d_{\alpha}$ the
\emph{dimension} of $A$. For example, every pure $n$-qubit
state expressed in a local basis, \be |\phi\rangle =
\sum_{i_1, \dots, i_n=0}^1 A_{i_1\dots i_n} |i_1\dots
i_n\rangle \ee corresponds to an $2\times\dots\times 2$
tensor of rank $n$ and dimension 2.

If  $A^{(1)}$ and $A^{(2)}$ are two tensors of ranks $n_1$
and $n_2$, respectively, and $s$ and $t$ are integers with
$1\leq s\leq n_1$ and $1\leq t\leq n_2$, and both the
$s^{\rm th}$ index of $A^{(1)}$ and the $t^{\rm th}$ index
of $A^{(2)}$ range from $1$ to the same integer $d$, then a
sum of the form \be \sum_{j=1}^d A^{(1)}_{i_1\dots i_{s-1}\
j\ i_{s+1}\dots i_n}A^{(2)}_{i_1\dots i_{t-1}\ j\
i_{t+1}\dots i_n}\ee yields a tensor of rank $n_1+n_2-1$.
This sum is called a \emph{contraction} of the tensors
$A^{(1)}$ and $A^{(2)}$. More specifically, one says that
the $s^{\rm th}$ index of $A^{(1)}$ is contracted with the
$t^{\rm th}$ index of $A^{(2)}$. A situation where several
tensors $A^{(1)}, \dots, A^{(N)}$ are contracted at various
indices is called a \emph{tensor network}.   The maximal
dimension of any tensor in the network, is called the
\emph{dimension} of the network, and will usually be
denoted by $D$ in the following. Note that every tensor
network with $n$ \emph{open} indices (i.e., indices which
are not contracted), can be associated in a natural way to
an $n$-party pure quantum state.

We will only consider tensor networks where every index
appears at most twice in the network. In this case, every
tensor network can be represented by a \emph{graph} $F$ in
the following way.

\begin{itemize}
\item For every tensor $A^{(\alpha)}$ a vertex $\alpha$ is
drawn. \item Whenever two tensors $A^{(\alpha)}$ and
$A^{(\beta)}$are contracted, an edge is drawn between the
corresponding vertices $\alpha$ and $\beta$ in the graph.
\item Finally, for every \emph{open} index of a tensor
$A^{(\alpha)}$, i.e., an index which is not contracted, one
draws a new vertex and an edge connecting this vertex to
the vertex $\alpha$.
\end{itemize}

\begin{figure}[ht]
\begin{picture}(230,120)
\put(-5,0){\epsfxsize=230pt\epsffile[27 337 992
745]{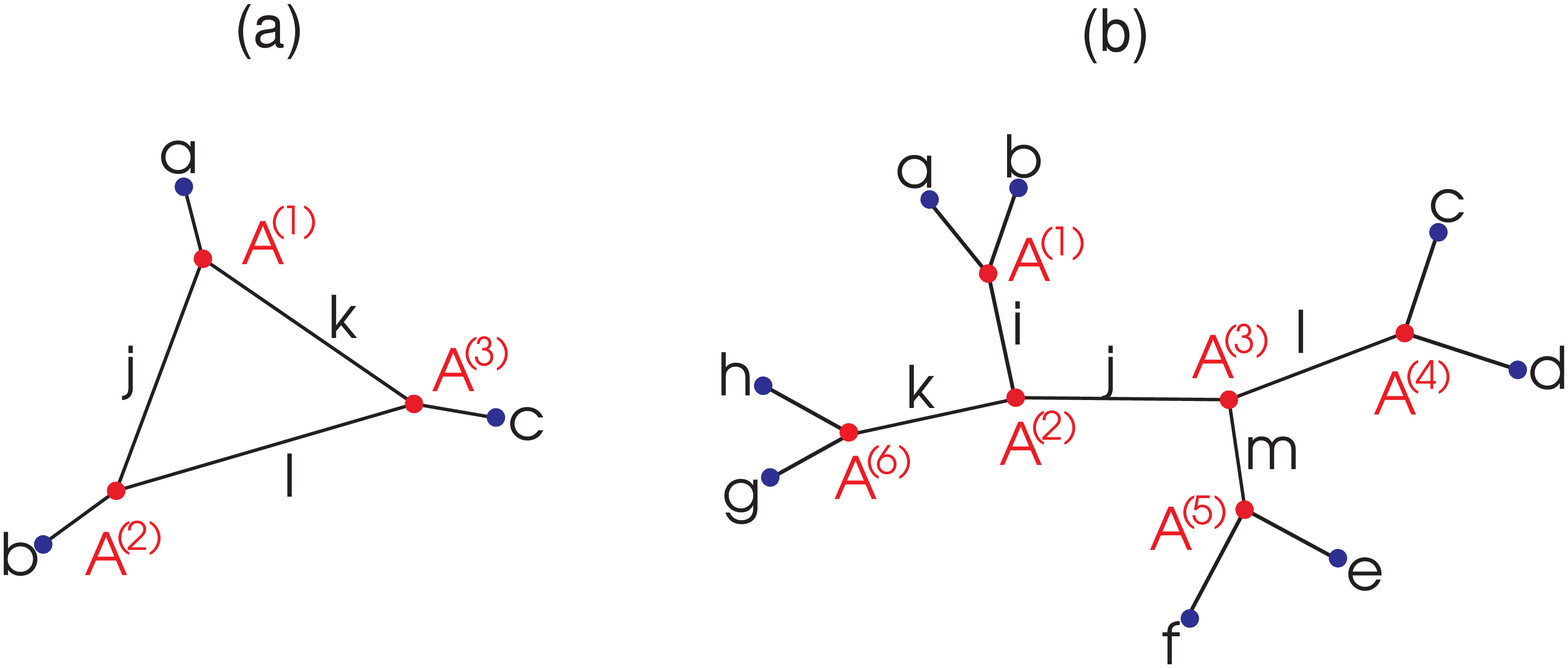}}
%\put(0,0){\includegraphics[width=0.45\textwidth]{tensor1.eps}}
\end{picture}
\caption[]{\label{Tensor1} Tensor network with three
tensors $A^{(1)}_{ajk},A^{(2)}_{bjl},A^{(3)}_{ckl}$ and
three open indices $a,b,c$ corresponding to a cycle graph.
(b) Tensor network with six tensors
$A^{(1)}_{abi},A^{(2)}_{ijk}, \ldots ,A^{(6)}_{ghk}$ and
eight open indices $a,b,c,d,e,f,g,h$ corresponding to a
tree graph.}
\end{figure}

As an example, consider three tensors $A^{(1)}, A^{(2)},
A^{(3)}$ contracted as follows: \be\label{TN}
\sum_{jkl}A^{(1)}_{ajk}A^{(2)}_{bjl}A^{(3)}_{ckl}.\ee This
tensor network has 3 open indices $a, b, c$,  and the
indices $j, k, l$ are contracted.  The graph underlying
this tensor network is depicted in Fig. \ref{Tensor1}a. The
tensor network (\ref{TN}) is naturally associated with a
$3$--partite pure state \be |\psi\rangle := \sum_{abc}
\left\{ \sum_{jkl}A^{(1)}_{ajk}A^{(2)}_{bjl}A^{(3)}_{ckl}
\right\}|a\rangle_1|b\rangle_2|c\rangle_3,\ee where we
introduced local bases $\{|a\rangle_1\}$,
$\{|b\rangle_2\}$, and $\{|c\rangle_3\}$ (the subscripts
denote the associated Hilbert spaces of the basis vectors).
In fact, $|\psi\rangle$ is an example of a matrix product
state. Writing \be |\psi_{jk}^{(1)}\rangle&:=&\sum_{a}
A^{(1)}_{ajk} |a\rangle_1,\nonumber\\
|\psi_{jl}^{(2)}\rangle&:=&\sum_{b} A^{(2)}_{bjl}
|b\rangle_2,\nonumber\\ |\psi_{kl}^{(3)}\rangle&:=&\sum_{c}
A^{(3)}_{ckl} |c\rangle_3,\nonumber\ee one obtains the
shorthand notation \be |\psi\rangle =
\sum_{jkl}|\psi_{jk}^{(1)}\rangle|\psi_{jl}^{(2)}\rangle|\psi_{kl}^{(3)}\rangle.\ee
It is clear that similar shorthand expressions can be
obtained for arbitrary tensor networks.

A \emph{tree tensor network} (TTN) is a tensor network
where the underlying graph is a tree, i.e., a graph with no
cycles. An example of a TTN is \be
\sum_{ijklm}
A^{(1)}_{abi}A^{(2)}_{ijk}A^{(3)}_{jlm}A^{(4)}_{cdl}A^{(5)}_{efm}A^{(6)}_{ghk},\ee
and the corresponding tree graph is depicted in Fig. \ref{Tensor1}b.
Note that (\ref{TN}) is an example of a tensor network
which is \emph{not} a TTN.

The following definitions regarding TTNs will be important
below (see theorem \ref{thm_TTN}). Let $T$ be a tree. An
\emph{open edge} is an edge which is incident with a leaf
of $T$. An \emph{inner edge} is an edge which is not an
open edge. Consider a TTN with tree $T$ having $n$ open
edges, corresponding to an $n$-party state $|\psi\rangle$.
Let $e\in T$ be an inner edge, and let $(A^e_T, B^e_T)$ be
the corresponding bipartition of the system. By
partitioning all tensors in the network in two classes as
induced by the bipartition $(A^e_T, B^e_T)$ and grouping
all contractions which occur between tensor in the same
class of the bipartition, one can write the network in the
form \be \sum_i
|\phi^i_{A^e_T}\rangle|\xi^i_{B^e_T}\rangle.\ee We say that
the TTN is in \emph{normal form w.r.t the bipartition
$(A^e_T, B^e_T)$} if the vectors
$\{|\phi^i_{A^e_T}\rangle\}$ and
$\{|\xi^i_{B^e_T}\rangle\}$ are (up to a normalization) the
Schmidt vectors of the state $|\psi\rangle$ w.r.t the
bipartition $(A^e_T, B^e_T)$. We say that the TTN is in
\emph{normal form} if it is in normal form for all
bipartitions $(A^e_T, B^e_T)$, where $e$ ranges over all
inner edges in $T$ \cite{Foot5}.

The interest in TTNs in quantum information theory  lies in
the property that the representation of systems in terms of
TTNs leads to efficient \emph{descriptions} of states as
well as to the possibility of efficiently simulating the
\emph{dynamics} of the system. The  main results in this
context were obtained in Refs. \cite{Sh05} and
\cite{Sh'05}. The latter result will be particularly
interesting for our purposes, and will be reviewed next.

We will be concerned with TTNs corresponding to subcubic
trees. It can easily be verified that if a TTN corresponds
to a subcubic tree, has $n$ open indices, and has dimension
$D$, then the TTN depends on at most $O(n D^{3})$ complex
parameters. Therefore, if an $n$-party state can be
described by a TTN where $D$ scales at most polynomially in
$n$, then $|\psi\rangle$ can be described by poly$(n)$
complex parameters by using this TTN. Hence a family of
systems allowing an efficient description is obtained. What
is more, it has been shown that also the \emph{processing}
of such systems can efficiently be simulated classically.
The following result, obtained in Ref. \cite{Sh'05}, will
play an important role in the subsequent analysis.

\begin{thm}\label{thm_shi}
 If an $n$-party pure quantum state $|\psi\rangle$ is
specified in terms of a TTN of dimension $D$, where the
underlying tree graph is subcubic, then any MQC performed
on $|\psi\rangle$ can be classically simulated in $O(n\
\mbox{poly(D)})$ time.
\end{thm} Therefore, if $D$ grows at most polynomially with
$n$, then the above simulation scheme is efficient. It is
noted by the authors in Ref. \cite{Sh'05} that there is no
restriction in considering subcubic trees only, in the
sense that any $n$-party state which can be represented by
a TTN (with arbitrary underlying tree) with poly$(n)$
parameters, can also be represented by a subcubic TTN with
poly$(n)$ parameters.

\subsection{Description of quantum systems with TTNs} \label{section_simTTN}

Theorem \ref{thm_shi} shows that, if an efficient TTN
description is \emph{known} for a quantum state, then LOCC
on this state can be simulated efficiently. However, this
result does not give any information about \emph{obtaining}
an (efficient) TTN description of a given state. Note that,
if a state is specified, there might exist several TTN
descriptions, some of which might be efficient and some of
which might not be. In fact, we will see below that, if a
subcubic tree with $n$ open edges is specified, then
\emph{any} $n$-party state $|\psi\rangle$ can be
represented by a TTN with  this specific tree structure --
although generally tensors of exponential dimension in $n$
are required. Therefore, the following two questions are
naturally raised:
\begin{itemize}
\item If a state $|\psi\rangle$ and a subcubic tree $T$ are
given, what is the behavior of the dimension $D$ of the
associated TTN(s)? \item If only a state $|\psi\rangle$ is
given, what is the optimal subcubic TTN describing this
state, i.e., the one with the smallest dimension $D$?
\end{itemize}
Next it is
shown that the entanglement in the state $|\psi\rangle$ as measured by the
Schmidt--rank, plays a crucial role in answering the above
questions. We prove the following result.

\begin{figure}[ht]
\begin{picture}(230,80)
\put(-5,0){\epsfxsize=230pt\epsffile[6 543 826
782]{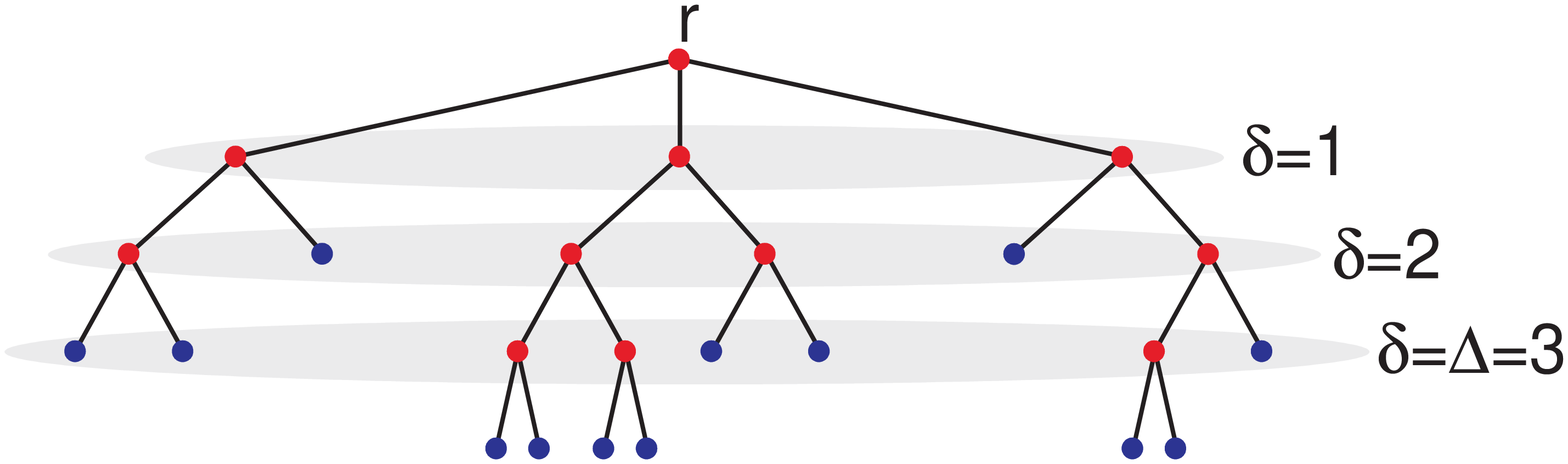}}
%\put(0,0){\includegraphics[width=0.45\textwidth]{tree1.eps}}
\end{picture}
\caption[]{\label{Tree1} Subcubic tree with root $r$, where
leaves (corresponding to the $n=13$ parties of the system)
are indicated in blue, and inner vertices are indicated in
red. The tree is arranged in such a way that all inner
vertices of same depth $\delta$ are on the same horizontal
line.}
\end{figure}

\begin{thm}\label{thm_TTN}
Let $|\psi\rangle$ be an $n$-party state and let $T$ be a
subcubic tree with $n$ leaves which are identified with the
$n$ parties in the system. Then there exists a TTN description
of $|\psi\rangle$ with underlying tree $T$, where the
dimension $D$ of this TTN is equal to \be\label{D} \log_2 D = \max_{e\in T}
\chi_{A^e_T, B^e_T}(|\psi\rangle).\ee Moreover, this TTN is in normal form.
\end{thm}
{\it Proof:}  the proof is constructive. The idea is to stepwise compute all tensors associated to the vertices of $T$, by traversing the tree from the leaves to the root, as depicted in Fig. \ref{Tree1}. First we need some definitions.  A vertex of $T$ which is not a leaf is called an \emph{inner} vertex; note that every inner vertex has degree 3. We fix one inner vertex $r$ and call it the \emph{root} of the tree $T$. The \emph{depth} of a vertex is the length of the shortest path from this vertex to the root $r$. We denote by $\Delta$ the maximal depth of any inner vertex in $T$. We refer to Fig. \ref{Tree1} for a schematic representation.

The construction is initialized  by considering all inner
vertices $\{v_1, \dots, v_N\}$ of depth $\Delta$. Every
such vertex has two open edges, corresponding to two qubits
in the system. We let $\{a_{\alpha}, b_{\alpha}\}$ be the
vertices associated in this way to $v_{\alpha}$, for every
$\alpha$.  We then compute all Schmidt decompositions
w.r.t. the bipartitions ($\{a_{\alpha}, b_{\alpha}\}$ --
rest of the system), i.e., \be\label{Schmidt_ini}
|\psi\rangle = \sum_{i}
|\phi_i^{(\alpha)}\rangle|\xi_i^{(\alpha)}\rangle, \ee for
every $\alpha$. The vectors $|\phi_i^{(\alpha)}\rangle$
have support on the qubits $\{a_{\alpha}, b_{\alpha}\}$,
the vectors $|\xi_i^{(\alpha)}\rangle$ have support on the
rest of the system. The Schmidt coefficients are absorbed
in the latter vectors.
%Writing () out with indices, we can read the r.h.s. of this expression, for every $\alpha$, as a tensor %network where a rank 3 tensor is contracted with a rank $(n-1)$ tensor at the index $i$. The rank 3 tensor %is $A^{(\alpha)}_{ijk}$, defined by  \be |\phi_i^{(\alpha)}\rangle:= \sum_{j, k}A^{(\alpha)}_{ijk} %|jk\rangle,\ee and will be asociated to the vertex $v_{\alpha}$. In this way one associates a rank 3 tensor %to every vertex $v_{\alpha}$ of depth $\Delta$, where every tensor has dimension smaller than (\ref{D}).

One then proceeds by computing  the tensors associated to
the inner vertices of depth $\Delta - 1$, and then to the
vertices of depth $\Delta - 2, \dots$, up to depth equal to
1, by in every step applying the procedure which will be
outlined now.

Let $1 \leq \delta\leq \Delta-1$.  For every vertex $v$,
let $T_v$ be the unique subtree of $T$ such that $v\in T_v$
and $T_v$ is one of the two subtrees obtained by deleting
the upper edge of $v$. Let $T_v^*$ be the tree obtained by,
first, adding one vertex $v^*$ to $T_v$ and connecting
$v^*$ to $v$ with an edge $\{v, v^*\}$ and, second, drawing
$\kappa$ open edges at the vertex $v^*$, where $\kappa$ is
equal to the number of qubits which do not correspond to
leaves of $T_v$.

Now, suppose that the following is  true: \emph{for all
inner vertices $w$ of depth $\delta+1$, a TTN description
for $|\psi\rangle$ is known with tree $T_w^*$, and all
these TTNs are in normal form.} We then outline a procedure
to obtain, for every inner vertex $v$ of depth $\delta$, a
TTN description for $|\psi\rangle$ with tree $T_v^*$, and
all these TTNs are in normal form.

\begin{figure}[ht]
\begin{picture}(230,310)
\put(-5,0){\epsfxsize=230pt\epsffile[10 8 736
945]{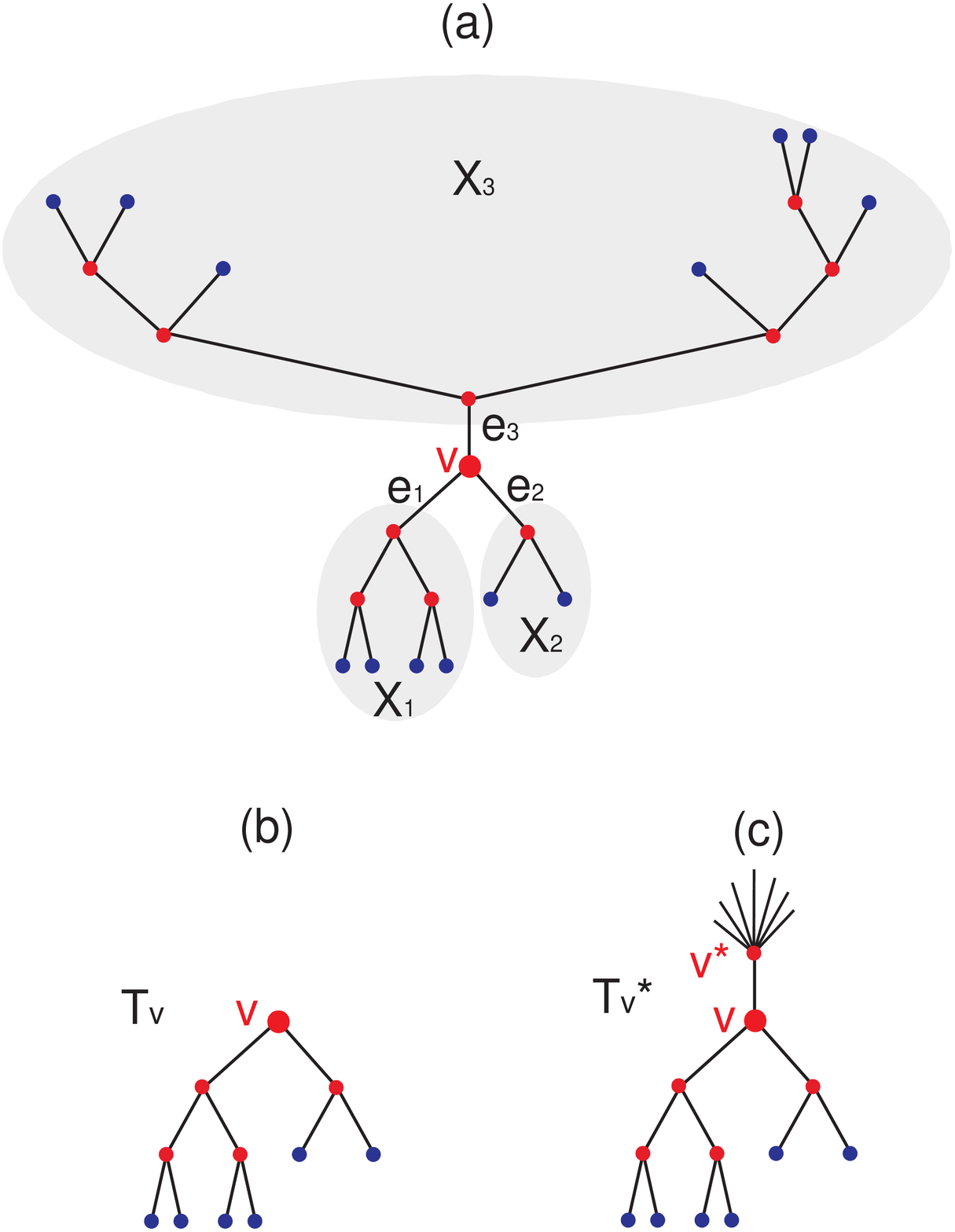}}
%\put(0,0){\includegraphics[width=0.45\textwidth]{tree2.eps}}
\end{picture}
\caption[]{\label{Tree2} (a) Same subcubic tree as depicted
in Fig. \ref{Tree1}, where vertices are re--arranged. We
consider an inner vertex $v$ of depth $\delta=1$ with lower
edges $e_1,e_2$ and upper edge $e_3$, and the corresponding
tripartition of the system into groups $X_1,X_2.X_3$. (b)
Subtree $T_v$ and (c) tree $T_v^*$ (for definition see
text).}
\end{figure}

{\bf Procedure.---} Consider an  inner vertex $v$ of depth
$\delta$. Let $e_1, e_2, e_3$ denote the edges incident
with $v$, such that $e_1$ and $e_2$ are the lower edges,
and $e_3$ is the upper edge as in Fig. \ref{Tree2}. Let
$(X_1, X_2, X_3)$ be the unique tripartition of the system
defined by \be (X_1, X_2\cup X_3) &:=& (A^{e_1}_T,
B^{e_1}_T)\nonumber\\ (X_2, X_1\cup X_3) &:=& (A^{e_2}_T,
B^{e_2}_T)\nonumber\\ (X_3, X_1\cup X_2) &:=& (A^{e_3}_T,
B^{e_3}_T).\ee See also Fig. \ref{Tree2} for a simple
pictorial definition.

We then make the distinction between the following cases:
\begin{itemize}
\item[(A)] neither $e_1$ or  $e_2$ are open edges, i.e.,
both edges connect $v$ to other inner vertices; \item[(B)]
one of these two edges, say $e_2$, is an open edge.
\end{itemize}
First we consider case (A). Let $v_1$ ($v_2$) be the vertex
connected to $v$ by the edge $e_1$ ($e_2$). By assumption,
we have TTN descriptions for $|\psi\rangle$ with trees
$T_{v_1}^*$ and $T_{v_2}^*$ which are in normal form.
Consider these TTN descriptions, and group all contractions
in such a way that one obtains Schmidt decompositions of
$|\psi\rangle$ with respect to the above bipartitions:
\be\label{Schmidt} |\psi\rangle =
\sum_{i=1}^{d_{\alpha}}|\psi^i_{X_i}\rangle|\psi^i_{\bar
X_i}\rangle, \ee for every $\alpha=1,2$, where
$d_{\alpha}:= \chi_{X_i, \bar X_i}(|\psi\rangle)$ denote
the Schmidt ranks, and where $\bar X_i$ denotes the
complement of $X_i$ (e.g., $\bar X_1 = X_2\cup X_3$). The
Schmidt coefficients have been absorbed in the vectors
$|\psi^i_{\bar X_i}\rangle$. Consider also the Schmidt
decomposition of $|\psi\rangle$ w.r.t. the split $(X_3,
X_1\cup X_2)$, using an analogous notation
\be\label{Schmidt2}|\psi\rangle =
\sum_{i=1}^{d_{3}}|\psi^i_{X_3}\rangle|\psi^i_{\bar
X_3}\rangle. \ee  The latter decomposition is not given by
TTN so far, and has to be calculated separately.   Using
the above 3 Schmidt decompositions, we can write \be
|\psi\rangle &=&
\sum_{i=1}^{d_{1}}|\psi^i_{X_1}\rangle|\psi^i_{X_2\cup
X_3}\rangle\label{1}\\ &=& \sum_{i=1}^{d_{1}} \
|\psi^i_{X_1}\rangle \langle\psi^i_{X_1}|\psi\rangle
\label{2}\\ &=&\sum_{i=1}^{d_{1}}\sum_{j=1}^{d_2}
|\psi^i_{X_1}\rangle |\psi^j_{X_2}\rangle
\langle\psi^i_{X_1}|\psi^j_{X_1\cup X_3}\rangle\label{3}\\
&=&  \sum_{i=1}^{d_{1}}\sum_{j=1}^{d_2}\sum_{k=1}^{d_3}
|\psi^i_{X_1}\rangle |\psi^j_{X_2}\rangle
|\psi^k_{X_3}\rangle B^{ijk},\label{4} \ee where we have
used the  following arguments and definitions. In order to
go from (\ref{2}) to (\ref{3}), we have inserted equation
(\ref{Schmidt}) for $\alpha=2$ in (\ref{2}); to obtain the
last equality (\ref{4}), we have defined the tensor
$B^{ijk}$ by \be\label{B}
\langle\psi^i_{X_1}|\psi^j_{X_1\cup X_3}\rangle=
\sum_{k=1}^{d_3} B^{ijk} |\psi^k_{X_3}\rangle. \ee This
yields a TTN description of $|\psi\rangle$ with underlying
tree $T_v^*$. Note that (\ref{4}) implies that the Schmidt
vectors $|\psi^k_{X_1\cup X_2}\rangle$ are recuperated as
\be|\psi^k_{X_1\cup X_2}\rangle = \sum_{i, j}
|\psi^i_{X_1}\rangle |\psi^j_{X_2}\rangle  B^{ijk},\ee
which shows that the TTN is in normal form w.r.t.the
bipartition $(X_3, X_1\cup X_2)$. It then immediately
follows that this TTN is in normal form. This concludes
case (A).

Next we consider case (B). Let $v_1$ and $v_2$ be  defined
as above. Note that in this case $X_2=\{v_2\}$. Consider
again the TTN description and related Schmidt decomposition
(\ref{Schmidt}) for $\alpha=1$, i.e., for the bipartition
$(X_1, \{v_2\}\cup X_3)$. Note that the Schmidt
decomposition for the split $(\{v_2\}, X_1\cup X_3)$ is not
available from the TTN since $v_2$ is not an inner vertex,
but we will not need it. As in (A), consider also the
Schmidt decomposition (\ref{Schmidt2}), i.e., for the
bipartition $(X_3, X_1\cup\{v_2\})$. We then write \be
|\psi\rangle &=& \sum_{i=1}^{d_{1}} \  |\psi^i_{X_1}\rangle
\langle\psi^i_{X_1}|\psi\rangle\\
&=&\sum_{i=1}^{d_{1}}\sum_{k=1}^{d_3} |\psi^i_{X_1}\rangle
|\psi^k_{X_3}\rangle \langle\psi^i_{X_1}|\psi^k_{X_1\cup
\{v_2\}}\rangle\\ &=& \sum_{i=1}^{d_{1}}\sum_{k=1}^{d_3}
|\psi^i_{X_1}\rangle  |\psi^{ik}_{\{v_2\}}\rangle
|\psi^k_{X_3}\rangle\ee  where we have used the definition
\be\label{psi_ik_v2} |\psi^{ik}_{\{v_2\}}\rangle:=
\langle\psi^i_{X_1}|\psi^k_{X_1\cup \{v_2\}}\rangle.\ee
This yields a TTN description of $|\psi\rangle$ with
underlying tree $T_v^*$ which is again in normal form. This
concludes (B). This also ends the procedure.

\

Note that the assumption of the procedure  is trivially
fulfilled for $\delta = \Delta -1$ after the Schmidt
decompositions (\ref{Schmidt_ini}) have been computed. The
procedure is then applied to $\delta = \Delta -1, \Delta
-2, \dots, 1$. After this, all tensors in the desired TTN
description are known, except the one associated to the
root $r$ of $T$. To obtain this final tensor, the following
steps are taken. Let $e_1$, $e_2$, $e_3$ be the edges
incident with $r$, let $v_1$, $v_2$, $v_3$ be the
corresponding vertices of depth 1, and let the tripartition
$(X_1, X_2, X_3)$ be defined as before. From the previous
steps in the algorithm, we have TTN descriptions for
$|\psi\rangle$ with trees $T_{v_1}^*$, $T_{v_2}^*$ and
$T_{v_3}^*$ which are in normal form. Consider these TTN
descriptions, and group all contractions as above,  in such
a way that one obtains Schmidt decompositions of
$|\psi\rangle$ with respect to the above bipartitions:
\be\label{Schmidt_final} |\psi\rangle =
\sum_{i=1}^{d_{\alpha}}|\psi^i_{X_i}\rangle|\psi^i_{\bar
X_i}\rangle, \ee for every $\alpha=1,2, 3$. A similar
derivation as (\ref{1})--(\ref{4}) shows that
$|\psi\rangle$ can be written as
\be\label{B'}\label{5}|\psi\rangle=
\sum_{i=1}^{d_{1}}\sum_{j=1}^{d_2}\sum_{k=1}^{d_3}
|\psi^i_{X_1}\rangle |\psi^j_{X_2}\rangle
|\psi^k_{X_3}\rangle B^{ijk},\ee where $B^{ijk}$ is defined
similarly as above. This expression describes
$|\psi\rangle$ as a TTN with tree $T$, as desired.
Moreover, it follows from (\ref{5}) that this TTN is in
normal form w.r.t to the bipartitions $(X_{\alpha}, \bar
X_{\alpha})$ for $\alpha=1,2,3$. Since the TTNs
(\ref{Schmidt_final}) were in normal form by construction,
this implies that the TTN description (\ref{5}) is in
normal form altogether.

Finally, it immediately follows that the dimension  of this
TTN is equal to (\ref{D}) \cite{Foot'}. This concludes the proof of
theorem \ref{thm_TTN}. \finpr

Note that theorem \ref{thm_TTN} proves that, if a subcubic
tree with $n$ open edges is specified, then any $n$-party
state can be represented by a TTN with  this specific tree
structure. The construction presented in the proof of
theorem \ref{thm_TTN}  is similar to a procedure presented
in Ref. \cite{Vi03} of how to obtain a matrix product
description (which is a particular instance of a tensor
network) for an arbitrary state $|\psi\rangle$; there, too,
the dimension of the tensor network depends on the maximal
Schmidt rank of $|\psi\rangle$ as measured w.r.t a specific
class of bipartite splits, similar to (but different from)
eq.  (\ref{D}).

\subsection{Connection with $\chi$--width}\label{sect_connection}

Theorem \ref{thm_TTN} now allows us to give a natural
interpretation of the $\chi$-width measure
(\ref{chiwidth}).  Namely, for any state $|\psi\rangle$ one
has:

\begin{itemize}
\item  $\chi_{\mbox{\scriptsize{wd}}} (|\psi\rangle)$ is
the  smallest possible dimension of a TTN associated to
$|\psi\rangle$ through the Schmidt decomposition
construction described in theorem \ref{thm_TTN}; \item the
tree $T$ which yields the minimum in (\ref{chiwidth})
corresponds to the optimal TTN, i.e., the one with smallest
dimension \cite{Foot6}.
\end{itemize}
These observations fully answer (Q1),  the first of the two
central questions put forward in section \ref{sect_problem}
of this article. What is more, we now immediately arrive at
a satisfactory answer to question (Q2), since theorems
\ref{thm_shi} and \ref{thm_TTN} (see also Ref.
\cite{Sh'05}) imply the following.
\begin{thm}\label{mainthm}
Let $|\psi\rangle$ be an $n$-party state. Denote
$\chi:=\chi_{\mbox{\scriptsize{wd}}}(|\psi\rangle)$, let
$T$ be a tree yielding the optimum in the definition of
$\chi$, and suppose that the TTN description of
$|\psi\rangle$ with underlying tree $T$ is known. Then any
MQC on $|\psi\rangle$ can be simulated classically in $O(n\
poly(2^{\chi}))$ time.
\end{thm}
In particular, this  result shows that, whenever
$\chi_{\mbox{\scriptsize{wd}}}$ is \emph{bounded} on a
family of states $\Psi=\{|\psi_1\rangle, |\psi_2\rangle,
\dots\}$, then any MQC on $\Psi$ can be simulated
efficiently classically -- even in linear time in the
system size $n$. This result fully answers question (Q2) in
the negative; i.e., the $\chi$--width measure, which was
originally introduced as a means to assess whether a
resource $\Psi$ is universal for MQC, can equally well be
used to asses whether MQC on $\Psi$ can be efficiently
simulated classically. In particular, we have found that
MQC can be simulated efficiently for \emph{any family
$\Psi$ which is ruled out by the $\chi$--width criterion
(i.e., theorem \ref{thm_ewd} (ii)) as being a
non--universal resource}.

Note that theorem \ref{mainthm} even allows one to conclude
that efficient simulation is possible when
$\chi_{\mbox{\scriptsize{wd}}}$ grows at most
logarithmically with the system size -- i.e., it may be
unbounded. One observes that if
$\chi_{\mbox{\scriptsize{wd}}}$ exhibits this scaling
behavior on a family of states $\Psi$, then it is not
detected by the $\chi$--width universality criterion. This
apparent paradox is resolved by considering the notion of
\emph{efficient universality}, which was briefly introduced
in section \ref{sect_uni}. When this requirement is
introduced in the definition of universality, the above
paradox is resolved as follows. One can prove \cite{Va06b}
that $\chi_{\mbox{\scriptsize{wd}}}$ (and
$E_{\mbox{\scriptsize{wd}}}$) need to grow \emph{faster
than logarithmically with the system size} on any efficient
universal resource. This clearly resolves the above
apparent contradiction.

While the above results indeed settle questions (Q1) and
(Q2), in practical situations one is of course faced with
the problem whether, when a state $|\psi\rangle$ is
specified, the optimal TTN can be computed efficiently. In
particular, if theorem \ref{mainthm} is to be applied, the
following quantities need to be computed:
\begin{itemize}
\item[(a)] The quantity $\chi$ itself; \item[(b)] an
optimal subcubic tree $T$ in the calculation of $\chi$;
\item[(c)] the TTN description of $|\psi\rangle$
corresponding to the tree $T$.
\end{itemize}
It is clear that, for any of the above quantities to be
efficiently computable, in the least one needs to have an
efficient description of the state $|\psi\rangle$ in some
form -- say, a polynomial size quantum circuit leading to
the preparation of the state, or, in the case where
$|\psi\rangle$ is a graph state, the underlying graph or
stabilizer description. If an efficient description is not
available, quantities such as e.g. the Schmidt rank w.r.t.
some bipartition can generally not be computed efficiently,
and there is no hope of computing e.g. (a) in polynomial
time. However, it is important to stress that the
possibility of an efficient description is by no means
sufficient to compute the quantities (a)--(b)--(c)
efficiently.

Regarding (a) and (b), the optimization in the definition
of the $\chi$--width measure suggests that an explicit
evaluation of $\chi_{\mbox{\scriptsize{wd}}}$ in a
specified state, as well as the determination of the
optimal subcubic tree, might be a highly nontrivial task.
However, we note that general results in this context are
known. In particular, we refer to Ref. \cite{Oum}, where
optimization problems of the form \be \min_{T}\max_{e\in T}
f(A^e_T)\ee are considered, where $f$ is a function defined
on subsets of $V:=\{1, \dots, n\}$, $f: A\subseteq V
\rightarrow f(A)$. It has been shown that such
optimizations can be performed in polynomial time in $n$,
i.e., the optimum as well as the tree yielding the optimum
can be determined efficiently, for a subclass of functions
$f$ which meet several technical requirements. In the next
section we will see that the graph states form a class of
states where these requirements are met, such that the
calculation of the $\chi$-width can be performed
efficiently.  However, the techniques presented in Ref.
\cite{Oum} might be used or generalized to calculate the
$\chi$--width efficiently for classes of states larger than
the graph states.

Regarding (c), it is clear that the optimal TTN description
of $|\psi\rangle$ can only be computed efficiently if this
TTN description is itself efficient, i.e., if it depends on
at most poly$(n)$ parameters -- this is exactly the case
when $\chi$ scales as log$(n)$. If $\chi$ scales in the
latter way, then it follows from the procedure outlined in
theorem \ref{thm_TTN}, that the optimal TTN description of
$|\psi\rangle$ can be obtained efficiently given one is
able to determine the following quantities in poly$(n)$
time:
\begin{itemize}
\item[(i)] the Schmidt coefficients and Schmidt vectors for
all bipartitions $(A^e_T, B^e_T)$, where $T$ is the optimal
tree in the definition of the $\chi$--width. \item[(ii)]
Certain overlaps between Schmidt vectors: in particular,
the tensor coefficients  \be B^{ijk} =
\langle\psi^i_{X_1}|\langle\psi^k_{X_3}|\psi^j_{X_1\cup
X_3}\rangle\ee in eq. (\ref{B}) and similar tensors in eq.
(\ref{B'}), as well as the vectors
\be|\psi^{ik}_{\{v_2\}}\rangle:=
\langle\psi^i_{X_1}|\psi^k_{X_1\cup \{v_2\}}\rangle\ee in
eq. (\ref{psi_ik_v2}).
\end{itemize}
Thus, a number of conditions need to be fulfilled to obtain
 an efficient TTN description, if it exists, for a given
state. Remarkably, in the next section we show that the
quantities (a), (b) and (c) can be computed efficiently for
all \emph{graph states}.

As a final remark in this section, note that an efficient
TTN description (if it exists) of a state $|\psi\rangle$
w.r.t. a given tree $T$, can always we obtained efficiently
if $|\psi\rangle$ is already specified in terms of an
efficient TTN description w.r.t. a different tree $T'$.

\section{Graph states}\label{sect_grapstates}

In this section we specialize the results obtained in the
previous section to graph states.

\subsection{Simulation of MQC}\label{sect_sim_graph}

Theorem \ref{mainthm} and the connection between
$\chi$--width of graph states and rank width of graphs,
allows us to obtain the following result.

\begin{thm}\label{mainthm2}
Let $|G\rangle$ be a graph state on $n$ qubits.  If the
rank width of $G$ grows at most logarithmically with $n$,
then any MQC on $|G\rangle$ can efficiently be simulated
classically.
\end{thm}
In particular,  the above result shows that if rwd$(G)$ is
bounded on  a family ${\cal G}=\{G_1, G_2, \dots\}$, then
any MQC on the set $\Psi({\cal G})=\{|G_1\rangle,
|G_2\rangle, \dots\}$ can efficiently be simulated
classically. This provides a complementary result to the
one obtained in Ref. \cite{Va06}, where it was proved that
any family of graphs with bounded rank width cannot provide
a universal resource for MQC. Therefore, all examples given
in Ref. \cite{Va06} of non--universal graph states (see
also section \ref{sect_uni}) can also be given here as
examples of resources on which MQC can be simulated
efficiently classically.

Note that theorem \ref{mainthm2} supersedes all known
results (see section \ref{sect_sim}) on classical
simulation of MQC on graph states. To see this, let us
consider the result in Ref. \cite{Sh05} stating that MQC
can be simulated efficiently on all graph states $G$ with
logarithmically bounded tree width twd$(G)$. Using the
inequality \cite{Ka06} \be \mbox{rwd}(G)\leq 4\cdot\mbox{
twd}(G) + 2,\ee one finds that, whenever twd$(G)$ scales as
log$(n)$ (where $n$ is the number of qubits in the system),
then also rwd$(G)$ scales at most as log$(n)$. Thus,
theorem \ref{mainthm2} implies that MQC can be simulated
efficiently on all graph states $G$ with logarithmically
bounded tree width, and the result in Ref. \cite{Sh05} is
retrieved. This shows that theorem \ref{mainthm2} fully
recovers and generalizes the known results on simulation of
MQC on graph states.

Finally, we emphasize that the rank width can be bounded on
families of graphs which do \emph{not at all} have any
tree--like structure, i.e., graphs possibly having many
cycles; therefore, the presence of cycles in a graph is no
indication that efficient simulation of MQC on the
associated  state might be hard. One reason of this
property is that a possible tree structure of a graph does
not remain invariant under local operations; e.g., the
fully connected graph and the star graph (one central
vertex connected to all other vertices) are locally
equivalent; the latter is a tree graph, the former is not
-- in fact, the tree width of the star graph is equal to 1,
whereas the tree width of the fully connected graph on $n$
vertices is $n-1$ \cite{Ro84}. Contrary to e.g. the
tree--width measure, the rank width is a local invariant,
thus taking into account such cases. Due to these
properties, our results prove a significant extension to
the use of the tree width; indeed, the above example
unambiguously illustrates the superiority of the rank width
as a criterion to address the classical simulation of MQC
on graph states.

\subsection{TTNs for graph states}

In this section we are concerned with the issue whether, if
a graph state is given, the optimal TTN can be computed
efficiently, i.e., we consider the quantities (a)--(b)--(c)
as denoted in section \ref{sect_connection}.

Let $G$ be a graph on $n$ vertices. It was shown in  Ref.
\cite{Oum} that, for a fixed integer $k$, the problem
\emph{"Is the rank width of $G$ smaller than $k$?"} is in
the complexity class $P$. Moreover, in Ref. \cite{Ou05}
several polynomial--time so--called \emph{approximation
algorithms} for the rank width are constructed. When $G$ is
given as an input, the (most efficient) algorithm either
confirms that rwd$(G)$ is larger than $k$, or it outputs a
subcubic tree $T^*$ such that \be \max_{e\in T^*} \mbox{
rank}_{\mathbb{F}_2} \Gamma( A_{T^*}^e, B_{T^*}^e) =
3k-1,\ee which implies that rwd$(G)\leq 3k-1$. The running
time of the algorithm is $O(n^3)$.

These results  immediately yield an efficient procedure to
determine the qualitative behavior of the $\chi$--width of
graph states, and to determine the optimal subcubic tree in
the calculation of the $\chi$--width. More precisely, a
possible (binary search) approach is the following: first
run the above algorithm for $k=n/2$; if the algorithm
confirms that rwd$(G)\geq n/2$, then run the algorithm for
$k=3n/4$; if not, then run the algorithm for $k=n/4$, etc.
This algorithm is guaranteed to terminate in poly$(n)$
time. After the last run of the algorithm, the rank width,
and the corresponding optimal subcubic tree, is obtained up
to a factor 3.

Thus, both quantities (a) and (b) as defined in the
discussion following theorem \ref{mainthm}, can be computed
efficiently for any graph state.

As for an efficient calculation of quantity (c), we note
that, for any bipartition of the system, both the Schmidt
coefficients and the Schmidt vectors can be computed
efficiently for graph states using the stabilizer
formalism; moreover, the Schmidt vectors can always be
chosen to be stabilizer states themselves. This can be
proved as follows (we only give a sketch of the argument,
as it involves standard stabilizer techniques). Let
$|G\rangle$ be a graph state on qubits $V:=\{1, \dots,
n\}$, and let $(A, B)$ be a bipartition of $V$. Let ${\cal
S}$ denote the stabilizer of $|G\rangle$, defined by
\be{\cal S}:= \left\{\prod_{a\in V} (K_a)^{x_a}\ |
x_a\in\{0, 1\}, \forall\ a\in V\right\},\ee where the
operators $K_a$ have been defined in eq. (\ref{K_a}). Thus,
${\cal S}$ is the commutative group generated by the
operators $K_a$. One then has \cite{He06} \be
|G\rangle\langle G| &=&\frac{1}{2^n} \prod_{a\in V}(I+
K_a)=l \frac{1}{2^n} \sum_{g\in{\cal S}} g.\ee Let ${\cal
S}_A$ be the subgroup of operators in ${\cal S}$ acting
trivially on the qubits in $V\setminus A = B$. Then
\be\rho_A:= \mbox{ Tr}_B |G\rangle\langle G| =
\frac{1}{2^{|A|}} \sum_{g\in {\cal S}_A} g.\ee This
operator satisfies \be(\rho_A)^2 &=& \frac{1}{2^{2|A|}}
\sum_{g\in{\cal S}_A} g \sum_{h\in{\cal S}_A}h\nonumber\\
&=& \frac{1}{2^{2|A|}}\sum_{g\in{\cal S}_A} \sum_{h\in{\cal
S}_A}h= \frac{|{\cal S}_A|}{2^{|A|}} \rho_A.\ee The second
equality holds since ${\cal S}_A$ is a group. Denoting $r:=
2^{|A|}|{\cal S}_A|^{-1}$, it follows that
$(r\rho_A)^2=r\rho_A,$ showing that $r\rho_A$ a projection
operator. Thus, all nonzero eigenvalues of this operator
are equal to 1. This shows that all nonzero eigenvalues of
$\rho_A$ (which are the squares of the Schmidt coefficients
of $|G\rangle$ w.r.t. the bipartition $(A, B)$) are equal
to $r^{-1}= 2^{-|A|}|{\cal S}_A|$. Moreover, as $\rho_A$
has unit trace, it follows that \be r^{-1}\cdot \mbox{rank}
(\rho_A) = 1, \ee such that the number of nonzero
eigenvalues of $\rho_A$ is equal to $r= 2^{|A|}|{\cal
S}_A|^{-1}$ \cite{Foot}.

The eigenvectors of $\rho_A$ can be computed as follows.
Let $\{K^A_1, \dots, K^A_s\}\subseteq {\cal S}_A$ denote a
minimal generating set of ${\cal S}_A$, where
$s:=\log_2|{\cal S}_A|$. Let $\{K^A_{s+1}, \dots,
K^A_{|A|}\}$ be additional Pauli operators, chosen in such
a way that \be\label{stab} \{K^A_1, \dots, K^A_s,
K^A_{s+1}, \dots, K^A_{|A|}\}\ee is a set of commuting and
independent operators; such a set always exists (though it
is non--unique) and can be computed efficiently, by using
the stabilizer formalism (see e.g. \cite{He06}). Note that
(\ref{stab}) is the generating set of a stabilizer state
$|\psi\rangle$ on the qubits in $A$, namely the state
\be|\psi\rangle\langle\psi| :=
\frac{1}{2^{|A|}}\prod_{i=1}^{|A|} (I + K^A_i) .\ee
Moreover, this state is an eigenstate of $\rho_A$. To see
this, note that $K^A_j |\psi\rangle\langle\psi| =
|\psi\rangle\langle\psi|$, and thus $K^A_j
|\psi\rangle=|\psi\rangle$, for every $j=1, \dots, s$. As
$\{K^A_j\}_{j=1}^s$ is a generating set of the group ${\cal
S}_A$, this last identity implies that
$g|\psi\rangle=|\psi\rangle$ for every $g\in {\cal S}_A$,
and therefore \be \rho_A|\psi\rangle&=&\frac{1}{2^{|A|}}
\sum_{g\in {\cal S}_A} g|\psi\rangle= \frac{|{\cal
S}_A|}{2^{|A|}}|\psi\rangle.\ee In order to obtain a basis
of eigenvectors, one considers the $2^{|A|- s} =
2^{|A|}|{\cal S}_A|^{-1}$ stabilizer states
$|\psi_{\alpha_{s+1}, \dots, \alpha_{|A|}}\rangle$ with
stabilizers generated by \be \{K^A_1, \dots, K^A_s,
\alpha_{s+1}K^A_{s+1}, \dots, \alpha_{|A|}K^A_{|A|}\}, \ee
where $\alpha_k=\pm 1$, for every $k=s+1, \dots, |A|$. One
can, with arguments analogous to above, show that all these
states are eigenvectors of $\rho_A$. Moreover, all these
states are mutually orthogonal; one has \be  &&
\langle\psi_{\alpha_{s+1}, \dots,
\alpha_{|A|}}|\psi_{\beta_{s+1}, \dots,
\beta_{|A|}}\rangle\quad \nonumber\\ &=&
(-1)^{\alpha_k}\langle\psi_{\alpha_{s+1}, \dots,
\alpha_{|A|}}|K^A_k|\psi_{\beta_{s+1}, \dots,
\beta_{|A|}}\rangle\nonumber\\ &=& (-1)^{\alpha_k +
\beta_k}\langle\psi_{\alpha_{s+1}, \dots,
\alpha_{|A|}}|\psi_{\beta_{s+1}, \dots,
\beta_{|A|}}\rangle,\label{orth},\ee for every $k=s+1,
\dots, |A|$,  where we have respectively used that
\be\langle\psi_{\alpha_{s+1}, \dots, \alpha_{|A|}}| =
(-1)^{\alpha_k}\langle\psi_{\alpha_{s+1}, \dots,
\alpha_{|A|}}|K^A_k \ee and \be K^A_k|\psi_{\beta_{s+1},
\dots, \beta_{|A|}}\rangle =
(-1)^{\beta_k}|\psi_{\beta_{s+1}, \dots,
\beta_{|A|}}\rangle.\ee It immediately follows from the
identity (\ref{orth}) that the states $|\psi_{\alpha_{s+1},
\dots, \alpha_{|A|}}\rangle$ are mutually orthogonal. Since
there are exactly $2^{|A|}|{\cal S}_A|^{-1}$ such vectors,
as many as there are nonzero Schmidt coefficients, we have
computed all Schmidt vectors of $|G\rangle$ w.r.t. the
bipartition $(A, B)$. Remark that at this point we only
have a stabilizer description of the Schmidt vectors; if
necessary, the expansion of these vectors in the
computational basis can be computed using the results in
Ref. \cite{De05}.

This shows that both Schmidt coefficients and Schmidt
vectors of $|G\rangle$ w.r.t. any bipartition $(A, B)$ can
be computed efficiently, and that the Schmidt vectors can
always be chosen to be stabilizer states. Moreover, note
that overlaps between stabilizer states can be computed
efficiently using stabilizer techniques, and we refer to
\cite{Aa04}, where this problem was considered.

Thus, all necessary ingredients (cf. (i)--(ii) in section
\ref{sect_connection}) needed for the efficient
construction of the optimal TTN of a graph state
$|G\rangle$, can be computed efficiently when rwd$(G)$
scales as log$(n)$.

We then arrive at the following result.

\begin{thm}
Let $|G\rangle$ be a graph state on $n$ qubits and denote
$\chi:=\chi_{\mbox{\scriptsize{wd}}}(|G\rangle)$. Then an
optimal subcubic tree $T$ in the definition of $\chi$ can
be computed in poly$(n)$ time. Moreover, if $\chi$ scales
as log$(n)$ then the TTN description of $|G\rangle$
corresponding to $T$ can be computed in poly$(n)$ time.
\end{thm}
Note that, in particular, the conditions of the above
theorem are fulfilled for all classes of graphs having
bounded rank width, and thus efficient TTNs can be computed
in poly$(n)$ time for all such classes.

\subsection{Example for the cycle graph on $n=6$ qubits}

In this section we give an explicit example of the
computation of the rank width, the optimal subcubic tree,
and the corresponding TTN description of a particular graph
state, namely the 6-qubit state $|C_6\rangle$ associated to
the \emph{cycle graph} (or \emph{ring graph}) $C_6$ on 6
vertices. The adjacency matrix $\Gamma$ of $C_6$ is the
$6\times 6$ matrix \be \left[ \begin{array}{cccccc}
\cdot&1&\cdot&\cdot&\cdot&1\\
1&\cdot&1&\cdot&\cdot&\cdot\\
\cdot&1&\cdot&1&\cdot&\cdot\\
\cdot&\cdot&1&\cdot&1&\cdot\\
\cdot&\cdot&\cdot&1&\cdot&1\\
1&\cdot&\cdot&\cdot&1&\cdot
\end{array}\right],\ee where ' $\cdot$ ' denotes an entry
equal to zero.

\subsubsection{Rank width and optimal tree}
First we compute the rank width of the graph $C_6$. In
fact, we will prove that rwd$(C_6)=2$. To show this,
consider the subcubic tree $T$ depicted in Fig.
\ref{subcubic}. The leaves of $T$ are associated to the
vertices of $C_6$ in the following natural way: first, fix
an arbitrary vertex of $C_6$ and denote this to be vertex
1; then, starting from vertex 1, traverse the vertices of
$C_6$ in a counterclockwise way, and denote the vertices by
2, 3, 4, 5 and 6, respectively; these vertices are now
associated to the leaves of $T$ by identifying vertex 1
with the leftmost leaf of $T$, vertex 2 with the second
leaf from the left, etc.

It is now straightforward to show that \be \max_{e\in T}
\mbox{ rank}_{\mathbb{F}_2}\ \Gamma(A^e_T, B^e_T) = 2. \ee
This can be showed by simply computing the ranks of all
matrices $\Gamma(A^e_T, B^e_T)$ and picking the largest of
these ranks. Furthermore, one has \be\label{alpha}
\alpha_{T'}(C_6):=\max_{e\in T'} \mbox{
rank}_{\mathbb{F}_2}\ \Gamma(A^e_{T'}, B^e_{T'}) \geq 2\ee
for every subcubic tree $T'$. This can be seen as follows:
first, note that $\alpha_{T'}(C_6)\geq 1$ for every $T'$,
since \be\mbox{rank}_{\mathbb{F}_2} \Gamma(A, B)\geq 1\ee
for every bipartition $(A, B)$. Second, suppose that $T'$
is a subcubic tree such that $\alpha_{T'}(C_6)= 1$; we will
show that this leads to a contradiction. Note that
rank$_{\mathbb{F}_2} \Gamma(A, B)$ is equal to 1  if and
only if $(A, B)$ is a bipartition of the form (one vertex
-- rest). Moreover, if $\alpha_{T'}(C_6)= 1$, then one must
have \be\mbox{ rank}_{\mathbb{F}_2}\ \Gamma(A^e_{T'},
B^e_{T'}) = 1\ee for every $e\in T'$. Thus, every
bipartition $(A^e_{T'}, B^e_{T'})$ must be of the form (one
vertex -- rest); this leads to a contradiction. This shows
that the inequality (\ref{alpha}) is correct. We can
therefore conclude that \be\mbox{rwd}(C_6):= \min_{T'}
\alpha_{T'}(C_6) = 2\ee and that the tree $T$ as depicted
in Fig. \ref{subcubic} yields the optimum.

At this point we note that here ad hoc methods have been
used to obtain the above result; however, we remind the
reader that general algorithms exist to calculate the rank
width and the optimal tree, as cited in section
\ref{sect_sim_graph}.

\subsubsection{TTN description}

The computation of the TTN description of $|C_6\rangle$
with underlying tree $T$ is performed in Appendix
\ref{App}. The result is the following:
\be\label{C_6_TTN'}\langle x_1 \dots x_6|C_6\rangle =
\frac{1}{2^3}\sum_{abcdef} \psi^{(1)}_{abx_1x_2}
\psi^{(2)}_{abcdx_3}\psi^{(3)}_{cdefx_4}\psi^{(4)}_{efx_5x_6},\nonumber\\
\ee where $x_1, \dots, x_6\in\{0, 1\}$ and where all
indices in the sum run from 0 to 1. The pair $ab$ should be
regarded as one index taking 4 different values, as well as
the pairs $cd$ and $ef$. Moreover, one has the following
definitions:\be \psi^{(1)}_{abx_1x_2}&:=&
\delta_{a ,x_1}\delta_{b, x_2}\nonumber\\
\psi^{(2)}_{abcdx_3}&:=&
(-1)^{ac+ ab + bx_3 + dx_3}\nonumber\\
\psi^{(3)}_{cdefx_4}&:=& \delta_{f,
c}\delta_{d, x_4}(-1)^{de + ec}\nonumber\\
\psi^{(4)}_{efx_5x_6}&:=& \delta_{e ,x_5}\delta_{f,
x_6}.\ee

\section{Complex systems versus tree structures in QIT and graph
theory}\label{sect_complex}

We have seen that the $\chi$--width of a graph state is
equal  to the rank width of the underlying graph. There is
in fact a striking parallel between the \emph{motivations}
for the definitions of rank width of graphs and of
$\chi$--width of general quantum states, on which we
comment here.

As explained above,  the $\chi$--width gives information
about the optimal TTN which describes a given quantum
state. The interest in such TTNs naturally arises due to
the fact that the dynamics of quantum systems which allow
TTN descriptions with sufficiently small dimension $D$, can
be simulated efficiently on a classical computer. These and
similar techniques (cf. e.g. the matrix product states
formalism) are invoked because the efficient classical
simulation of \emph{general} quantum systems can be a very
difficult problem. Thus, in spite of the general hardness
of this simulation problem, it becomes tractable when
restricted to the class of those systems with efficient TTN
descriptions.

In graph theory an  analogous situation occurs. While many
interesting problems are hard to compute on general graphs,
they become tractable for those classes of graphs which can
be associated, through certain constructions, with tree
structures. The simplest examples are of course the tree
graphs themselves, which are in some sense the simplest
instances of graphs; and indeed, many difficult problems
become efficiently solvable, or even trivial, on trees.
However, this is far from the whole story. In graph theory
one has considered a variety of so--called \emph{width
parameters}, which all measure, in different ways, how
similar a graph is to a tree graph. Examples are rank
width, tree width, clique--width, path--width, and
branch--width. It has been shown that for families of
graphs where a given width parameter is \emph{bounded},
large classes of (NP--)hard problems have efficient
solutions. For example, the problem of deciding whether a
graph is 3--colorable, which is a NP--hard, is efficiently
solvable when restricted to classes of graphs of bounded
rank width. The graph theoretical results in this context
are often very general and far--reaching; e.g., it has been
show that all graph problems which can be formulated in
terms of a certain mathematical logic calculus, have
efficient solutions when restricted to graphs of bounded
rank width. We refer to Ref. \cite{Hi06} for an accessible
treatment of these and related issues.

Thus, in certain aspects of both quantum information theory
and graph theory there is a natural interest in using tree
structures for the approximation of complex systems.
Moreover, there seems to be a strong parallel in the
explicit constructions which are used in both fields. A
striking example is obtained here,  as the rank width of
graphs exactly coincides with the $\chi$--width measure on
graph states. As a second example, it was found in Ref.
\cite{Sh05} that the efficient contraction of large tensor
network is directly related to the tree width of the
underlying graphs. The present authors believe that the
aforementioned parallel can significantly be exploited
further.

\section{Conclusion}\label{sect_conclusion}

In this paper we have considered the possibility to
classically simulate measurement based quantum computation.
We have shown that all states with a bounded or
logarithmically growing Schmidt--rank width can in fact be
described efficiently, and moreover any one--way quantum
computation performed on such states can also be simulated
efficiently. We have given an interpretation of the
Schmidt--rank width, a measure that has its origin in graph
theory, in terms of the optimal tree tensor network
describing a state. We have also provided a constructive
procedure how to obtain the optimal TTN, and discussed the
requirements that this can be done efficiently. For graph
states, we have explicitly constructed the corresponding
TTN, and provided an efficient algorithm to do this for any
graph state where the underlying graph has bounded or
logarithmically growing rank width. These results on
efficient simulation complement recent findings on
universality of states, in the sense that all states that
are found to be non--universal resources for MQC using the
Schmidt--rank width criteria (i.e. which have bounded
Schmidt--rank width) can also be simulated efficiently on a
classical computer. The connection to complexity issues in
graph theory, also highlighted in this paper, seems to
provide future possibilities for a fruitful interchange of
concepts and methods between the fields of quantum
information and graph theory.

\begin{acknowledgements}
This work was supported by the FWF, the European Union (QICS, OLAQUI, SCALA), and the \"OAW through project APART (W.D.).
\end{acknowledgements}

\appendix
\section{Optimal TTN description of $|C_6\rangle$}\label{App}

We now compute the TTN description of $|C_6\rangle$ w.r.t.
the tree $T$ depicted in Fig. \ref{subcubic}, using the
procedure outlined in theorem \ref{thm_TTN}. Consider the
following Schmidt decompositions of $|C_6\rangle$:
\be\label{C_6_Schmidt} |C_6\rangle
&=&\frac{1}{\chi^{(1)}}\sum_{i} |\phi^{(1)}_i\rangle_{12}
|\xi^{(1)}_i\rangle_{3456} \label{12_3456}\\
&=&\frac{1}{\chi^{(2)}}\sum_{j} |\phi^{(2)}_j\rangle_{123}
|\xi^{(2)}_j\rangle_{456} \label{123_456}\\
&=&\frac{1}{\chi^{(3)}}\sum_{k} |\phi^{(3)}_k\rangle_{1234}
|\xi^{(3)}_k\rangle_{56}. \label{1234_56}\ee These decompositions are
taken w.r.t. the bipartitions $(\{1, 2\}, \{3, 4, 5, 6\})$,
$(\{1, 2, 3\}, \{4, 5, 6\})$ and $(\{1, 2, 3, 4\}, \{5,
6\})$, respectively; these correspond to the bipartitions
$(A^e_T, B^e_T)$, where $e$ runs over all inner edges of
$T$.
All Schmidt vectors in (\ref{C_6_Schmidt}) are
normalized, and the $\chi^{(\alpha)}$ are the square roots
of the Schmidt ranks of the corresponding bipartitions
\cite{Foot7}.

We now show how the TTN description of $|C_6\rangle$ w.r.t
the tree $T$ is obtained,  by applying the procedure
presented in theorem \ref{thm_TTN}. First, note that the
depth $\Delta$ of $T$ is equal to 3. We start by
considering the single inner vertex of depth 3; this is the
vertex which has leaves 1 and 2 as lower vertices. We then
compute the Schmidt decomposition (\ref{12_3456}),
corresponding to the bipartition which is obtained by
deleting the upper edge of this vertex. In a second step,
we consider the single vertex in $T$ having depth 2, and
compute the corresponding Schmidt decomposition
(\ref{123_456}). Moreover, we write \be |C_6\rangle &=&
\frac{1}{\chi^{(1)}}\sum_{i} |\phi^{(1)}_i\rangle
\langle\phi^{(1)}_i|C_6\rangle\nonumber\\ &=&
\frac{1}{\chi^{(1)}\chi^{(2)}}\sum_{i,j}
|\phi^{(1)}_i\rangle
\langle\phi^{(1)}_i|\phi^{(2)}_j\rangle
|\xi^{(2)}_j\rangle,\label{intermediate}\ee (where we have
omitted the subscripts of the Schmidt vectors). Finally, we
consider the Schmidt decomposition (\ref{1234_56})
(corresponding to the uper edge of the unique depth 1
vertex), and write it as \be |C_6\rangle
&=&\frac{1}{\chi^{(3)}}\sum_{k}
|\xi^{(3)}_k\rangle\langle\xi^{(3)}_k|C_6\rangle\label{1234_56'}\ee
Combining eqs. (\ref{1234_56'}) and (\ref{intermediate})
then shows that $|C_6\rangle$ can be written as follows
\cite{Foot8}: \be\label{C_6_TTN}|C_6\rangle =
\frac{1}{\chi^{(1)}\chi^{(2)}\chi^{(3)}}\sum_{ijk}
|\phi^{(1)}_i\rangle \langle \phi^{(1)}_i|
\phi^{(2)}_j\rangle |\xi^{(3)}_k\rangle \langle\xi^{(3)}_k|
\xi^{(2)}_j\rangle.\nonumber\\\ee  Note that the states
$\langle \phi^{(1)}_i| \phi^{(2)}_j\rangle$ are defined on
qubit 3, for every $i$ and $j$, and that the states
$\langle\xi^{(3)}_k| \xi^{(2)}_j\rangle$ are defined on
qubit 4, for every $j$ and $k$.

Next we explicitly compute the Schmidt coefficients and
Schmidt vectors in the above expansions, using the
stabilizer formalism.

As for the Schmidt coefficients, note that \be 2 &=&\mbox{
rank}_{\mathbb{F}_2}\ \Gamma(\{1, 2\}, \{3, 4, 5, 6\})\nonumber\\
&=&\mbox{
rank}_{\mathbb{F}_2}\ \Gamma(\{1, 2, 3\}, \{4, 5, 6\})\nonumber\\
&=&\mbox{ rank}_{\mathbb{F}_2}\ \Gamma(\{1, 2, 3, 4\}, \{5,
6\}), \ee and therefore (using (\ref{ctrk})) all the
Schmidt ranks of the above bipartitions are equal to
$2^2=4$. Thus, the indices $i, j, k$ in eq. (\ref{C_6_TTN})
all run from 1 to $4$, and we also have \be \chi^{(1)}
=\chi^{(2)} = \chi^{(3)} = \sqrt{4}=2.\ee It will be
convenient to write the indices $i, j, k$ as pairs of bits,
and we will use the notations $i\equiv ab$, $j\equiv cd$,
$k\equiv ef$, where $a, b, c, d\in\{0, 1\}$.

We now consider the Schmidt vectors w.r.t. the above
bipartitions. We start with the bipartition $(\{1, 2\},
\{3, 4, 5, 6\})$. Here one finds that
\be\mbox{Tr}_{\{3,4,5,6\}} (|C_6\rangle\langle C_6|) =
\frac{1}{4}I.\ee Thus, a Schmidt basis for the subset $\{1,
2\}$ could simply be chosen to be the computational basis;
in other words, we take \be |\phi^{(1)}_{ab}\rangle =
|a\rangle\otimes |b\rangle \equiv |ab\rangle, \ee defined
on the qubits $\{1, 2\}$, for every $a, b\in\{0, 1\}$ .

The same argument can be repeated for the vectors
$\{|\xi^{(3)}_{ef}\}$, where we can take $
|\xi^{(3)}_{ef}\rangle = |ef\rangle$, defined on the qubits
$\{5, 6\}$, for every $e, f\in\{0, 1\}$.

As for the bipartition $(\{1, 2, 3\}, \{4, 5, 6\})$, one
can easily show that \be \mbox{Tr}_{\{4, 5, 6\}}
(|C_6\rangle\langle C_6|) = \frac{1}{8}(I + \sigma_z\otimes
\sigma_x\otimes \sigma_z)\ee and that, hence, the states
\be\label{LC}  |\phi^{(2)}_{cd}\rangle=\sigma_z^c\otimes
I\otimes\sigma_z^d|L_3\rangle\ee form a valid Schmidt
basis, where $c, d\in\{0, 1\}$ and where $|L_3\rangle$ is
the linear cluster state on $3$ qubits, defined on the
qubits $\{1, 2, 3\}$.

To compute the vectors $\{| \xi^{(2)}_{cd}\rangle\}$, note
that one has \be | \xi^{(2)}_{cd}\rangle = 2\langle
\phi^{(2)}_{cd}|C_6\rangle.\ee Therefore, we have to
compute expressions of the form \be\label{overlap0}
[(\langle L_3| \sigma_z^c\otimes I\otimes\sigma_z^d)\otimes
I] |C_6\rangle,\ee for every $c, d=0, 1$. To do so, we use
that every $n$-qubit graph state $|G\rangle$ with adjacency
$\Gamma$ can be written as
\cite{He06}\be\label{exp}|G\rangle =
\frac{1}{2^{n/2}}\sum_{u\in\{0, 1\}^n}
(-1)^{q_G(u)}|u\rangle, \ee where $\{|u\rangle\ |\ u\in\{0,
1\}^n\}$ is the $n$-qubit computational basis and where \be
q_G(u):= \frac{1}{2} u^T\Gamma u.\ee One then finds that
(\ref{overlap0}) is equal to (omitting multiplicative
constants) \be\label{overlap0'} \sum_{u, v, w}\left(
 \sum_{x, y, z} (-1)^{q_{C_6}(x, y, z, u, v, w)
+ q_{L_3}(x, y, z) + xc +
zd}\right)|uvw\rangle.\nonumber\\\ee Straightforward
algebra then shows that the power of $-1$ in the above
expression is equal to \be x(w+c) + z(d+u) + q_{L_3}(u, v,
w).\ee Moreover, one has \be \sum_{x, y, z} (-1)^{x(w+c) +
z(d+u)} = \left\{
\begin{array}{cl} 2^3& w=c \mbox{ and } d=u\\ 0&
\mbox{else}\end{array}\right.\ee We then find that
(\ref{overlap0'}) is equal to \be
|d\rangle\left(\sum_{v=0}^1(-1)^{q_{L_3}(d,v,c)}|v\rangle
\right) |c\rangle,\ee for every $c, d=0, 1$. Thus, these 4
states form the set $\{| \xi^{(2)}_{cd}\rangle\}_{j=1}^4$,
defined on qubits $\{4, 5, 6\}$.

The only remaining task is the computation of the states
$\langle \phi^{(1)}_{ab}| \phi^{(2)}_{cd}\rangle$ and
$\langle\xi^{(3)}_{ef}| \xi^{(2)}_{cd}\rangle$. To compute
the former of these states, it follows from the above that
one has to compute, for every $a, b, c, d\in\{0, 1\}$,
overlaps of the form \be\label{overlap1}\langle
\phi^{(1)}_{ab}| \phi^{(2)}_{cd}\rangle&=&\left(\langle
a|\otimes \langle b| \otimes I \right)\left(
\sigma_z^c\otimes I\otimes\sigma_z^d|L_3\rangle\right)\nonumber\\
&=& (-1)^{ac} \langle a|\otimes \langle
b|\otimes\sigma_z^d|L_3\rangle. \ee  Using the expansion
(\ref{exp}), it is then easy to show that (\ref{overlap1})
is equal to \be (-1)^{ac}\sum_{v=0}^1 (-1)^{q_{L_3}(a, b,
v)+dv} |v\rangle, \ee for every $a, b, c, d\in\{0, 1\}$,
and these states are defined on qubit 3. A similar
calculation can be performed to obtain \be \langle
\xi^{(3)}_{ef}| \xi^{(2)}_{cd}\rangle=\delta_{f,
c}(-1)^{q_{L_3}(d, e, c)}|d\rangle,\ee for every $ c, d, e,
f\in\{0, 1\}$, and these states are defined on qubit 4.

We can now write down the TTN description of the state
$|C_6\rangle$ w.r.t. the tree $T$ depicted in Fig.
\ref{subcubic}:

\be |C_6\rangle =
\frac{1}{2^{3}}\sum_{abcdef}\left\{|ab\rangle_{12} \left(
\sum_{v}(-1)^{ac+ q_{L_3}(a, b, v) +
dv}|v\rangle_3\right)\times\right.\nonumber\\ \left.\left(
\delta_{f, c}(-1)^{q_{L_3}(d, e, c)}|d\rangle_4\right)
|ef\rangle_{56}\right\},\nonumber\\\ee where we have again
indicated subscripts to specify on which qubits the states
are defined.

Recalling the definition of  $q_{L_3}$, namely \be
q_{L_3}(t_1, t_2, t_3):= t_1t_2 + t_2 t_3,\ee for every
$t_1, t_2, t_3\in\{0, 1\}$, we recover expression
(\ref{C_6_TTN'}). Note that one can easily check that
(\ref{C_6_TTN'}) is correct, by summing out all indices $a,
\dots, f$: \be\langle x_1 \dots x_6|C_6\rangle &=&
\frac{1}{2^3}\sum_{abcdef} \left\{\delta_{a ,x_1}\delta_{b,
x_2} (-1)^{ac+ ab + bx_3 + dx_3}\right. \nonumber\\&&
\left.\delta_{f, c}\delta_{d, x_4}(-1)^{de + ec}\delta_{e
,x_5}\delta_{f, x_6}\right\}\nonumber\\ &=& \frac{1}{2^3}
(-1)^{x_6x_1+x_1x_2+\dots + x_5x_6} \nonumber\\&=&
\frac{1}{2^3} (-1)^{q_{C_6}(x_1, \dots, x_6)},\ee where in
the last equality we indeed obtain the correct computational basis
expansion of $|C_6\rangle$.


\begin{thebibliography}{99}
\bibitem{Foot1}
We note that essential ingredients to the proof of this
result are already contained in earlier work \cite{Sh'05}.
We refer to section \ref{sect_tensor} for a precise
formulation of results.
\bibitem{Foot3}
The tree width is a graph parameter which measures how
similar a graph is to a tree. The tree width of any tree is
equal to 1 \cite{Ro84}.
\bibitem{Foot4}
The reader is referred to Ref. \cite{Ch68} for extensive
material regarding complex tensors.
\bibitem{Foot5}
The normal form defined here is closely related to the
canonical form of a TTN as defined in Ref. \cite{Sh'05}.
\bibitem{Foot'}
To be precise, we have only considered bipartitions w.r.t. \emph{inner} edges of the tree, whereas in (\ref{D}) all edges are considered. However, bipartitions w.r.t open edges always have the form (1 party -- rest of the system), which in any reasonable situation can be disregarded.
\bibitem{Foot6}
Note that there might exist several optimal trees.
\bibitem{Foot}
Moreover, one can show (see e.g. \cite{He06})
that $\log_2 r = \mbox{ rank}_{\mathbb{F}_2}\Gamma(A, B)$,
as used in eq. (\ref{ctrk}).
\bibitem{Foot7}
Recall that the nonzero Schmidt coefficients of
any bipartition of a system in a stabilizer state are equal
to the inverse of the square root of the Schmidt rank.
\bibitem{Foot8}
Note that in this case there is no need to treat the root
in the tree $T$ separately. This is because, in this
specific case, the root of $T$ has 2 open edges and only
one inner edge.
\end{thebibliography}
\end{document}